\documentclass{aa}
\usepackage{txfonts}
\usepackage{graphicx}
\usepackage{amssymb}
\usepackage{CJKutf8}
\usepackage{natbib}

\usepackage{scalerel}
\usepackage{tikz}
\usetikzlibrary{svg.path}

\usepackage{xcolor}
\usepackage{url}
\newcommand{\rom}[1]{\uppercase\expandafter{\romannumeral #1\relax}}

\bibpunct{(}{)}{;}{a}{}{,} 

\begin{document}

   \title{Reconstructing AGN X-ray spectral parameter distributions with Bayesian methods. \rom{1}: Spectral analysis}

   \author{Lingsong Ge
          \begin{CJK*}{UTF8}{gkai}
          (葛泠松)
          \end{CJK*}
          \inst{1}
          \and
          St\'ephane Paltani\inst{1}
          \and
          Dominique Eckert\inst{1}
          }

   \institute{Department of Astronomy,  University of Geneva,
              ch. d'Écogia 16, CH-1290 Versoix, Switzerland\\
              \email{lingsong.ge@unige.ch}
             }

   \date{Received xx, 2021; accepted xx, 202x}

  \abstract{X-ray spectra of active galactic nuclei (AGN) consist of several different emission and absorption components. To determine the spectral parameters, these components are often fitted manually with models chosen on a case-by-case basis. However, this approach has two problems. First, it becomes very hard for a survey with a large number of sources. Second, when the signal-to-noise ratio (S/N) is low, there is a tendency to adopt an overly simplistic model, biasing the parameters and making their uncertainties unrealistic. We developed a Bayesian method for automatically fitting AGN X-ray spectra obtained by \emph{XMM-Newton} with a consistent and physically motivated model. Our model includes all spectral components, even when the data quality is low. We used a physical model for the X-ray background and an empirical model for the non-X-ray background. Noninformative priors were applied on the parameters of interest, the photon index ($\Gamma$) and the hydrogen column density ($N_\mathrm{H}$), while informative priors obtained from deep surveys were used to marginalize over the parameter space of the nuisance parameters. To improve speed, we developed a specific spectral extraction and fitting procedure. We tested this method using a realistic sample of 5000 spectra, which was simulated based on our source model, reproducing typical population properties. Spectral parameters were randomly drawn from the priors, taking the luminosity function into account. Well-constrained or meaningful posterior probability density distributions (PDFs) were obtained for the most relevant spectral parameters, for instance, $N_\mathrm{H}$, $\Gamma,$ and $L_\mathrm{X}$ , even at low S/N, but in this case, we were unable to constrain the parameters of secondary components such as the reflection and soft excess. As a comparison, a maximum-likelihood approach with model selection among six models of different complexities was also applied to this sample. We find clear failures in the measurement of $\Gamma$ in most cases, and of $N_\mathrm{H}$ when the source is unabsorbed ($N_\mathrm{H} < 10^{22}$\,cm$^{-2}$). The results can hardly be used to reconstruct the parent distributions of the spectral parameters, while our Bayesian method provides meaningful multidimensional posteriors that will be used in a subsequent paper to infer the population.}

   
   \titlerunning{Reconstructing AGN X-ray spectral parameter distributions (\rom{1})}
   \authorrunning{L. Ge et al.}

   \maketitle

\section{Introduction}

Active galactic nuclei (AGN) are thought to be powered by accretion of galactic matter onto supermassive black holes (SMBHs) in the centers of galaxies. Extremely strong radiation is produced from the accretion power, rivaling and sometimes outshining the collective emission of all the stars in their host galaxy. The tremendous amount of energy released by AGN may also influence their surroundings, namely the host galaxies or galaxy clusters \citep[for a review, see][]{Fabian2012}. The co-evolution of AGN and host galaxies was first discovered in the tight correlation between the SMBH masses and the properties of the host galaxy bulges, such as the luminosity \citep{Magorrian1998} or velocity dispersion \citep{Gebhardt2000,Ferrarese2000}, and was then further detected in the similar histories of AGN accretion and star formation, which both peak at a redshift around 1--2 \citep{Hopkins2006, Silverman2008, Aird2015}. However, the details of the interplay between AGN and galaxies are still unclear \citep{Kormendy2013}. Therefore, much effort has been devoted to studying the demographics of AGN properties, such as the luminosity, accretion rate, and black hole mass, which can be used to reconstruct the mass accretion history of SMBHs in the Universe.

A reliable measurement of the cosmological build-up of SMBHs requires a complete census of AGN activities. However, this is very difficult to conduct due to the multifaceted AGN radiation properties at all wavelengths, which result from the large number of physical processes that produce the different components of the total AGN emission, and from the presence of obscuring material in the vicinity of AGN that strongly affects the direct observations by absorbing and reprocessing the primary radiation. A large fraction of AGN (up to about 70\%, depending on luminosity) is obscured by the intervening gas and dust \citep{Burlon2011,Ueda2014,Aird2015,Buchner2015,Ricci2015,Ananna2019}; it is therefore very important to characterize the full distribution of the absorbing column density and its dependence on redshift and AGN luminosity to uncover the complete AGN population.

Each method of AGN detection and selection has different biases \citep{Hickox2018}. Among them, X-ray surveys are particularly efficient for three main reasons: first, X-ray emission is ubiquitous in AGN \citep{Avni1986,Brandt2000,Gibson2008}; second, X-rays are able to penetrate a large amount of obscuring gas and dust along the line of sight because the photoelectric cross-section decreases with increasing photon energy \citep{Wilms2000}; and last, X-ray emission from the host galaxies is usually much weaker than that of the AGN, therefore X-rays are less diluted by stellar light than optical or IR surveys, for instance \citep{Maccacaro1988}.

 X-ray spectra of AGN are intrinsically very complex, including the X-ray power-law continuum, which is predominately produced by the inverse Compton scattering of disk UV/optical photons in a hot corona very close to the SMBH \citep{Blandford1990,Zdziarski1995,Zdziarski1996,Krolik1999}, an excess of soft X-ray emission below 1--2\,keV \citep{Arnaud1985,Singh1985}, the reflection of the primary X-ray power law by circumnuclear material \citep{Pounds1990,Nandra1994}, the scattering by warm fully ionized gas, located at a relatively large distance to the center of AGN \citep{Bianchi2006,Guainazzi2007}, and thermal emission from diffuse plasma in the host galaxy \citep{Iwasawa2011}. Therefore, they may show different characteristics in different cases, such as strong reflection in absorbed spectra or obvious soft excess in unabsorbed ones.

However, these features are very often difficult to detect or constrain in a wide survey; in a typical survey, most sources have low signal-to-noise ratios (S/N), so that we are most likely to detect only the main components. To cope with this difficulty, the classical approach attempts to adapt the chosen spectral model to the quality of the spectra at hand. One starts with the simplest model, and then visually inspects the residuals to assess if more components are needed. For example, \citet{Ricci2017} used more than 20 models to fit a set of about 800 individual sources. However, this approach is hardly applicable to survey data including a large number sources. For instance, the XXL survey has about 22\,000 X-ray point sources \citep{Pierre2016}, and the eROSITA all-sky survey \citep[eRASS;][]{Merloni2012} will produce a sample of about three million X-ray selected AGN \citep{Kolodzigr2013}. Moreover, in the case of low S/N, even an overly simplistic model may achieve statistically satisfactory fits \citep{Tozzi2006, Corral2011}. Adopting an unrealistic model may introduce biases and provide unrealistic errors in recovering the main parameters, for example, hydrogen column density ($N_\mathrm{H}$), photon index ($\Gamma$), and X-ray luminosity ($L_{\mathrm{X}}$). An example to illustrate this point is shown in Sect.~\ref{bias example}. Therefore, a consistent, physically motivated model and an automatic fitting procedure are required to fit AGN spectra that are extracted from large X-ray surveys.

In this paper we develop a new approach to analyzing the AGN X-ray spectral properties based on Bayesian inference, which allows us to properly propagate the uncertainties of components that are difficult to measure by marginalizing over them. The posteriors are computed with a nested-sampling algorithm to correctly deal with multimodal posteriors. Our goal is to devise a method that is capable of fitting the AGN X-ray spectra from about 0.5 to 12\,keV automatically with a consistent, physically motivated model without the need of model comparison and selection. At the same time, the method must be able to determine the most relevant spectral parameters ($N_\mathrm{H}$, $\Gamma$, and $L_{\mathrm{X}}$) without strong bias even at low S/N. The final goal of our study is to reconstruct the parent distributions of the spectral parameters from their posteriors, which is addressed in \citet{Ge2021} (submitted; hereafter Paper~\rom{2}). Throughout the paper, we consider the case of \emph{XMM-Newton} \citep{Jansen2001}, but the principles can be adapted to any other X-ray instrument with adequate changes.

\section{AGN X-ray spectral modeling}
\label{spectral modelling} 

\subsection{Components of X-ray emissions in AGN}
\label{source model}

At least five components contribute to the total X-ray spectra of AGN: the primary emission, soft excess, reflection, scattering in a warm plasma, and emission from thermal plasma in the galaxy. The primary emission is usually the dominant component, while the others are less luminous and more difficult to detect. In addition, the emitted spectrum is usually absorbed by gas and dust along the line of sight, both within the AGN and in our own galaxy. The detectability of the components depends strongly on the distribution of circumnuclear material in the AGN, which affects reflection and scattering.

\subsubsection{Primary emission}
The intrinsic X-ray emission from AGN is thought to be due to optical and UV photons from the accretion disk that are inverse-Compton scattered by high-energy electrons in a hot (kT$\sim$100\,keV), optically thin ($\tau\sim$0.5) corona \citep{Blandford1990, Zdziarski1995, Zdziarski1996, Krolik1999} in the vicinity of the SMBH. Some recent works suggest that the corona can be optically thick, although this depends on the assumed geometry \citep{Lanzuisi2019,Balokovic2020}. We fit this component using a redshifted power-law component with a high-energy cutoff with  \texttt{zcutoffpl} \citep[\texttt{xspec} model;][]{Arnaud1996}. Here the cutoff energy was fixed arbitrarily to 300\,keV \citep{Balokovic2020}, but because we only considered X-ray emission below $\sim$10\,keV the chosen energy is irrelevant. The free parameters of this component are the photon index ($\Gamma$) and the normalization ($F$), which is defined as the intrinsic flux of the cutoff power law in the 2 to 10\,keV band.

\subsubsection{Soft excess}
An excess of X-ray emission below 1--2\,keV, called soft excess, is widely detected in AGN X-ray spectra \citep{Piconcelli2005, Bianchi2009, Scott2012}, although its origin remains debated. Possible explanations include blurred ionization reflection \citep{Crummy2006, Fabian2009} or inverse Comptonization of disk photons in a warm, optically thick plasma \citep{Mehdipour2011, Done2012, Boissay2014}. Regardless of the origin of this component, we modeled it using a redshifted Bremsstrahlung (\texttt{zbrem}), which provides a simple phenomenological representation of its smooth spectral shape. In addition, the same shape has been used in \citet{Boissay2016}, which provides us with a distribution of the soft-excess strength in nearby sources. The soft-excess strength is defined as the flux ratio of the Bremsstrahlung in the 0.5--2\,keV band and the primary power law in the 2--10\,keV band \citep{Boissay2016}. We parameterized the soft-excess strength with a multiplicative constant ($q$), which  links the normalization of \texttt{zbrem} to the normalization parameter of \texttt{zcutoffpl} $F$. \citet{Ricci2017} used another phenomenological model with a black body to model the soft-excess component, and obtained a good measurement of its temperature: \(0.110 \pm 0.003\)\,keV. We therefore fixed the temperature in our \texttt{zbrem} model at 0.204\,keV, which gives the best fit to a black body at 0.110\,keV in the considered energy range. $q$ is then the only free parameter of the soft excess.

\subsubsection{Reflection}
The primary X-ray emission from the inner region of the AGN can be reprocessed by reflection on circumnuclear material, including the disk, broad-line region, and torus \citep{Pounds1990,Nandra1994}. We used \texttt{pexmon} \citep{Nandra2007} to model the reflection component, assuming a disk geometry and reflection with self-consistently generated fluorescence lines $\mathrm{Fe}$\,$\mathrm{K_\alpha}$, $\mathrm{Fe}$\,$\mathrm{K_\beta}$ , and $\mathrm{Ni}$\,$\mathrm{K_\alpha}$, as well as the $\mathrm{Fe}$\,$\mathrm{K_\alpha}$ Compton shoulder \citep{Nandra2007}. The true reflection spectrum depends on the precise shape and gas content of the reflector, but the expectation is that reflection will peak above 10\,keV in all reasonable geometries and densities. Constraining the true shape of the reflection component would require deep hard X-ray observations, which cannot be done with \emph{XMM-Newton} alone. Hence, \texttt{pexmon} was used as a model-independent description of the reflected component. The photon index, normalization, and cutoff energy were linked to those of the primary power law, and the inclination angle and the metallicity were fixed to $60^\circ $ and 1, respectively. Therefore the only free parameter of the reflection is the reflection fraction ($R$) relative to an infinite disk. 

\subsubsection{Scattering}
Primary power law and soft-excess emission escaping from the central engine can be scattered by fully ionized, optically thin gas relatively far away from the center of the AGN \citep{Bianchi2006,Guainazzi2007}. The fully ionized plasma acts as a mirror, and the shape of the scattered component remains essentially unchanged. Its flux is typically a few percent ($<10\%$) of the incoming radiation, however, because the gas is optically thin and intercepts only a small fraction of the line of sights \citep{Ueda2007}. We added a multiplicative constant $f_\mathrm{scatter}$ to the sum of the scattered \texttt{zcutoffpl} and \texttt{zbrem} without absorption from the torus to renormalize the flux. $f_\mathrm{scatter}$ is the only free parameter of this component. 

\subsubsection{Absorption}
Intrinsic absorption of the X-ray radiation is caused by two effects: the photoelectric absorption modeled with \texttt{zphabs,} and Compton scattering modeled with \texttt{cabs}. We linked the hydrogen column density $N_\mathrm{H}$ of these two absorption models, which is the only free parameter. The absorption only affects emission from the inner region of the AGN, therefore it was applied on the primary power law and on the soft excess alone.

\subsubsection{Thermal emission}
Thermal emission coming from diffuse plasma in the host galaxy of the AGN \citep{Iwasawa2011}, originating, for example, from star-forming regions, is modeled with a collisionally ionized plasma (\texttt{apec}). The free parameters were the temperature of the plasma (\textit{kT}) and its intrinsic luminosity \textit{lumin} in the 2--10\,keV band. The default normalization parameter was converted into luminosity using the convolution model \texttt{clumin}.

\subsubsection{Full model}
\label{full model}
The total AGN emission was then further absorbed by the gas in our galaxy. We modeled this absorption using \texttt{TBabs} \citep{Wilms2000}. Compared to \texttt{phabs}, which contains only the gas component, \texttt{TBabs} also includes dust, so it is a better model for absorption in our own galaxy. For a real survey, Galactic $N_\mathrm{H}$ is fixed for each source to the value at its sky coordinates, for instance, using the HI4PI Map \citep{Bekhti2016}. In the end, the full model, which has eight free parameters, is expressed (in \texttt{xspec} notation) as follows:

\texttt{TBabs $\times$ (zphabs $\times$ cabs $\times$ (zcutoffpl\\ + $q \times$ zbremss)  + clumin $\times$ apec + pexmon\\ + $f_\mathrm{scatter}$ $\times$ (zcutoffpl + $q \times$ zbremss)).}

\subsection{Classic spectral modeling}
\label{real example}

The complexity of X-ray spectra of AGN, as demonstrated by the number of free parameters, requires high data quality to fully characterize all the components. However, wide surveys designed to observe a large number of AGN to study their population properties are dominated in number by faint sources with modest S/N. For instance, more than 90\% of the sources in the XXL survey have an S/N lower than 3 \citep[3XLSS catalogue; ][]{Chiappetti2018}. Consequently, the full physical model is too complex for most of the spectra obtained in wide surveys. 

\begin{figure}[tb]
\centering
\resizebox{\hsize}{!}{\includegraphics{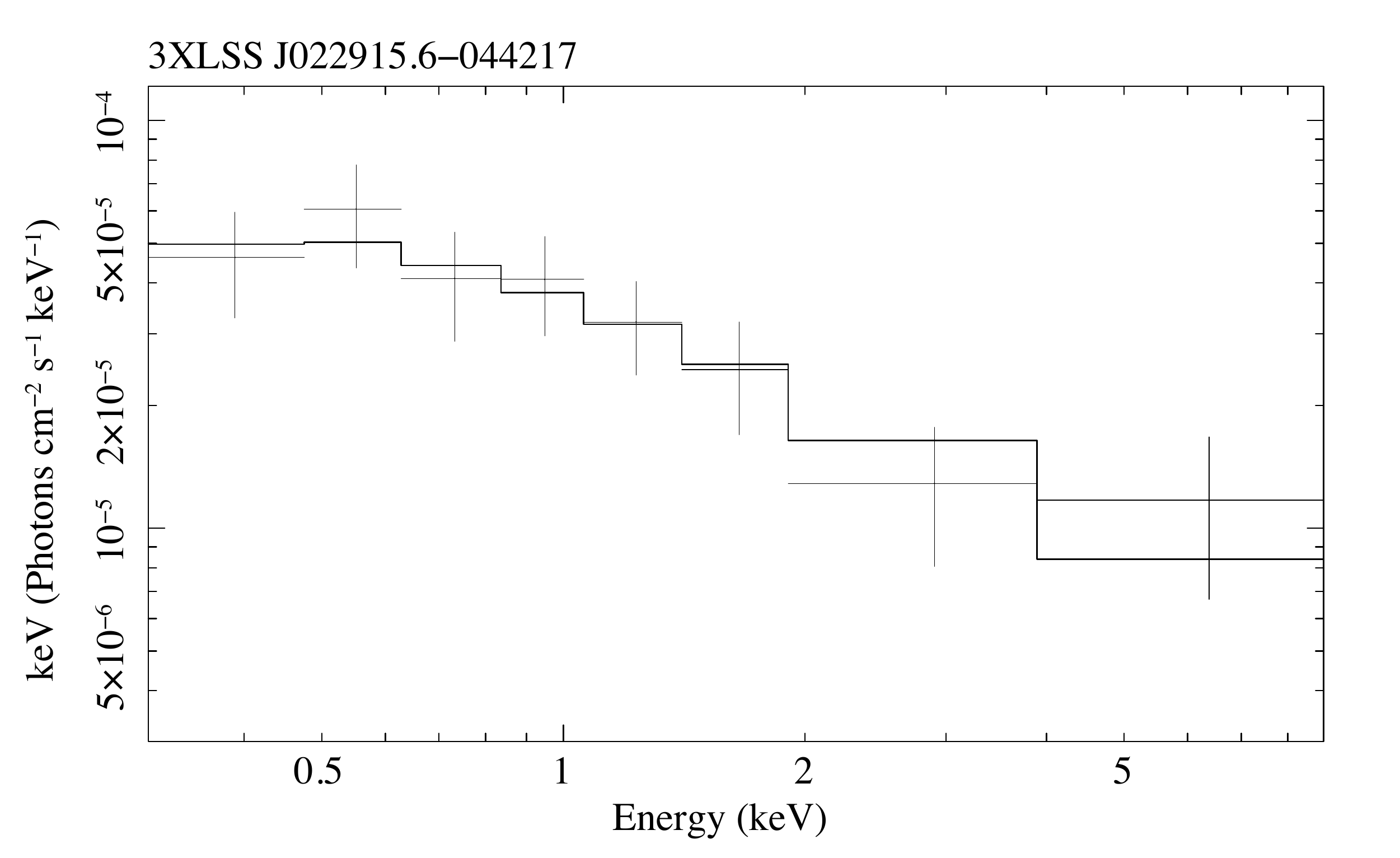}}
\resizebox{\hsize}{!}{\includegraphics{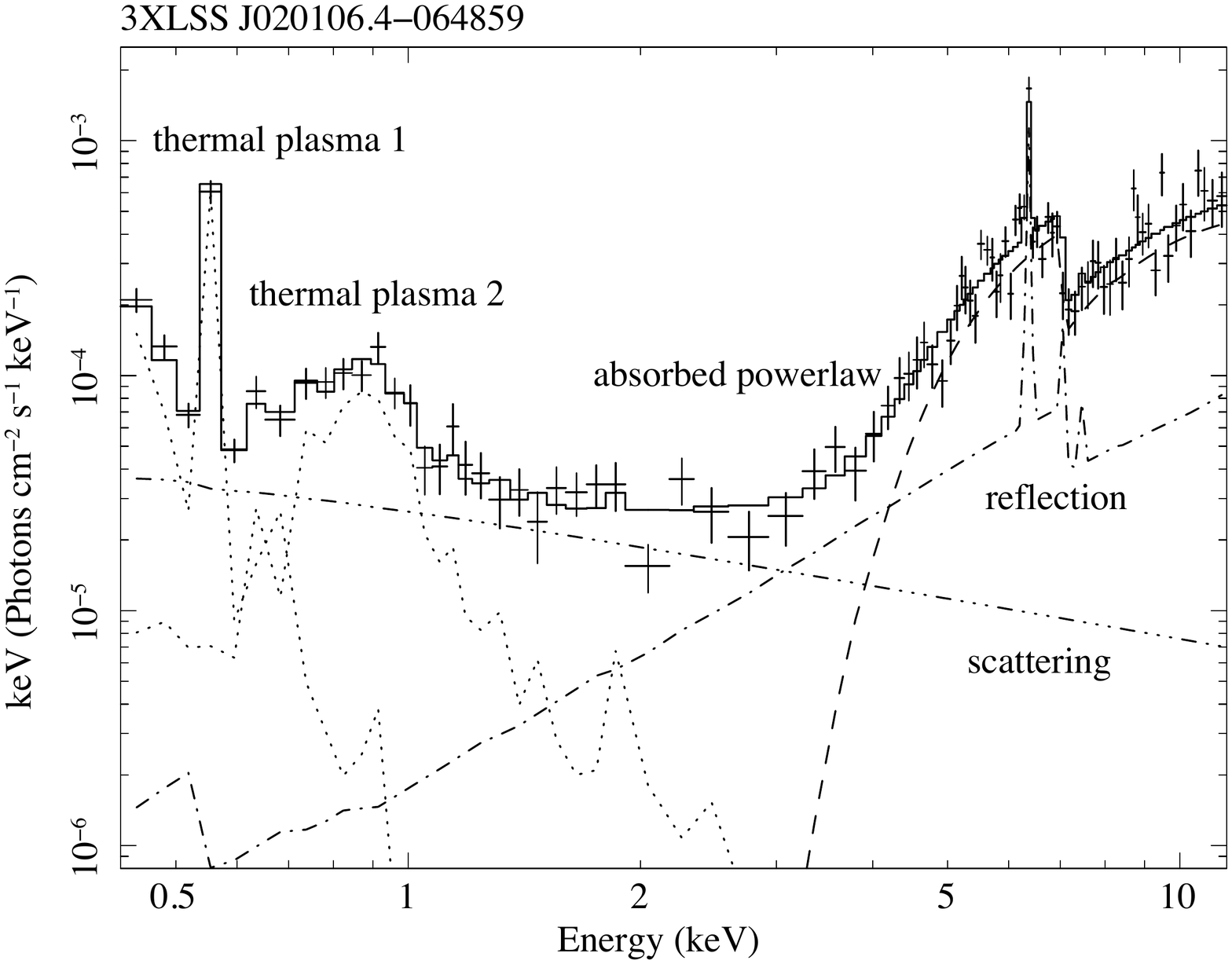}}
\caption{Unfolded energy spectra (\texttt{eufs}) of two sources from the XXL-1000-AGN sample \citep{Fotopoulou2016} that show very different spectral complexities. Top panel: 3XLSS J022915.6-044217. Bottom panel: 3XLSS J020106.4-064859 (NGC\,788). The simple spectrum can be fit with a simple power-law model, while the complex spectrum requires the full model. In the bottom panel different components of the spectrum are shown by dashed lines. They are identified with text.}
\label{fig:realfit}
\end{figure}

A common practice in the study of AGN X-ray spectra is to optimize some statistics, for instance, $\chi^2$ or C-stat \citep{Cash1979}, that is linked to the likelihood of a given parameter set. The best-fit parameters are then obtained using a maximum-likelihood (ML) optimization. However, when the number of free parameters is large compared to the S/N, the likelihood function is relatively flat over a large fraction of the parameter space, such that the position of the likelihood peak is sensitive to statistical fluctuations, making the ML optimization unstable.

To address the conflict between complexity of the model and low quality of the data, the model is usually simplified to include only the main components. Depending on the quality and characteristics of the spectrum, more complex models can be adopted. For example, we selected two sources from the 1000 brightest point source catalog of the XXL survey \citep{Fotopoulou2016} and fit them using \texttt{xspec}. The first source (3XLSS J022915.6-044217) has a simple spectrum with a medium to high S/N of 4.8 and an exposure time of 5.4\,ks. When we take the pn spectrum as an example, which has 122 spectral data counts and at least five counts per bin, it can be well fit using a simple power-law model with acceptable statistics (32.27 $\chi^2$ and 29.42 C-stat using 39 bins), as shown in Fig.~\ref{fig:realfit}. The data do not require more parameters, so that parameters such as $\Gamma$ and $N_\mathrm{H}$ are obtained from this simple fit. The second spectrum, from the Seyfert 2 galaxy \object{NGC\,788} (3XLSS J020106.4-064859, exposure time 10.0\,ks), has a much higher S/N (41) and shows many more spectral features, for example, a highly absorbed primary power law and a strong reflection with an iron emission line, as shown in the bottom panel of Fig.~\ref{fig:realfit}. This object is the brightest point source in the XXL north field \citep{Chiappetti2018}, with a 2--10\,keV flux of $1.4 \times 10^{-12}$ erg\,s$^{-1}$\,cm$^{-2}$. For this source, it is necessary to apply the full model (Sect.\,\ref{full model}), and each free parameter can be well fit, with the exception of the soft excess, which is completely absorbed\footnote{This object even requires two \texttt{apec} at different temperatures because its S/N is very high.}.

Although it is able to describe  the spectra of the sources of interest adequately, the model selection approach faces one important problem: a specific model for each source needs to be chosen. One can start with the simplest model and increase the model complexity depending on the improvement in the fit, as determined from an F-test, for instance, until the spectrum is well modeled. However, in a survey in which most sources have moderate fluxes, most sources are fitted with simple models \citep{Tozzi2006, Corral2011}. Consequently, the contributing but insignificant components are often ignored, resulting in a biased inference of the parameters (see an example in Sect.~\ref{bias example}).

\subsection{Bayesian approach}
\label{Bayesian appraoch}

Instead of the classical approach described above, which requires model selection, a Bayesian X-ray spectral analysis approach can be adopted. The main differences with the ML approach are first, that each parameter is conditioned by prior knowledge, and second, that the result of the fitting process is a fully multivariate distribution, namely the posterior, and not just a best solution with errors.

A Bayesian model of the X-ray spectra consists of the determination of the posterior probability $P(\Theta|y)$, with $\Theta$ being the set of parameters and $y$ the X-ray spectrum, using Bayes' theorem,

\begin{equation}
P(\Theta|y) = \frac{P(\Theta) \cdot P(y|\Theta)}{P(y)}
\label{eq:bayes}
.\end{equation}
where $P(\Theta)$ is the prior probability of the set of parameters $\Theta$, and $P(y|\Theta)$ is the likelihood of the observed data assuming the parameter set $\Theta$. $P(y)$ is the model evidence and can be ignored because all integrated probabilities give 1. We can then marginalize over the parameters of the less important components (i.e., sum the probabilities over all possible values of these parameters, e.g., all parameters except for $\Gamma$, $N_\mathrm{H}$ , and $F$) to propagate the uncertainty on the exact spectral shape of the source to the posterior distributions of the parameters of interest, even if they have little effect on the likelihood,

\begin{equation}
P(\Gamma, N_\mathrm{H}, F|y) = \int_{\Theta-\{\Gamma, N_\mathrm{H}, F\}} \frac{P(\Theta) \cdot P(y|\Theta)}{P(y)} d(\Theta-\{\Gamma, N_\mathrm{H}, F\})
\label{eq:marginalization}
.\end{equation}

The Bayesian approach introduced here means that model selection is no longer required as long as meaningful priors are used, and we can apply a consistent model to all the spectra. This avoids the potential biases that may be introduced by an overly simplistic model (see Sect.~\ref{bias example} below).

\subsection{Effect of the secondary components}
\label{bias example}

\begin{figure*}[tb]
\centering
\includegraphics[width=\columnwidth]{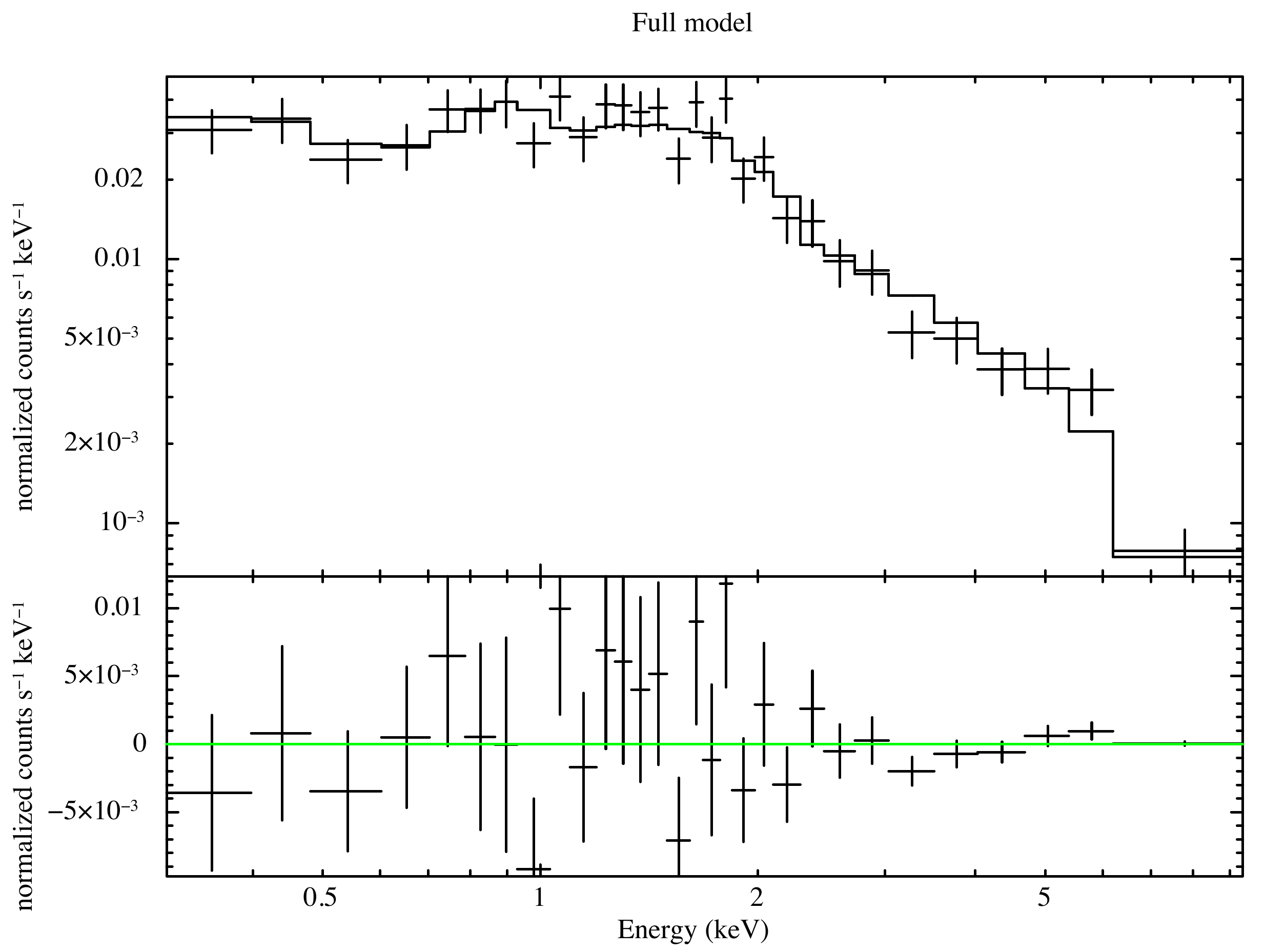}
\includegraphics[width=\columnwidth]{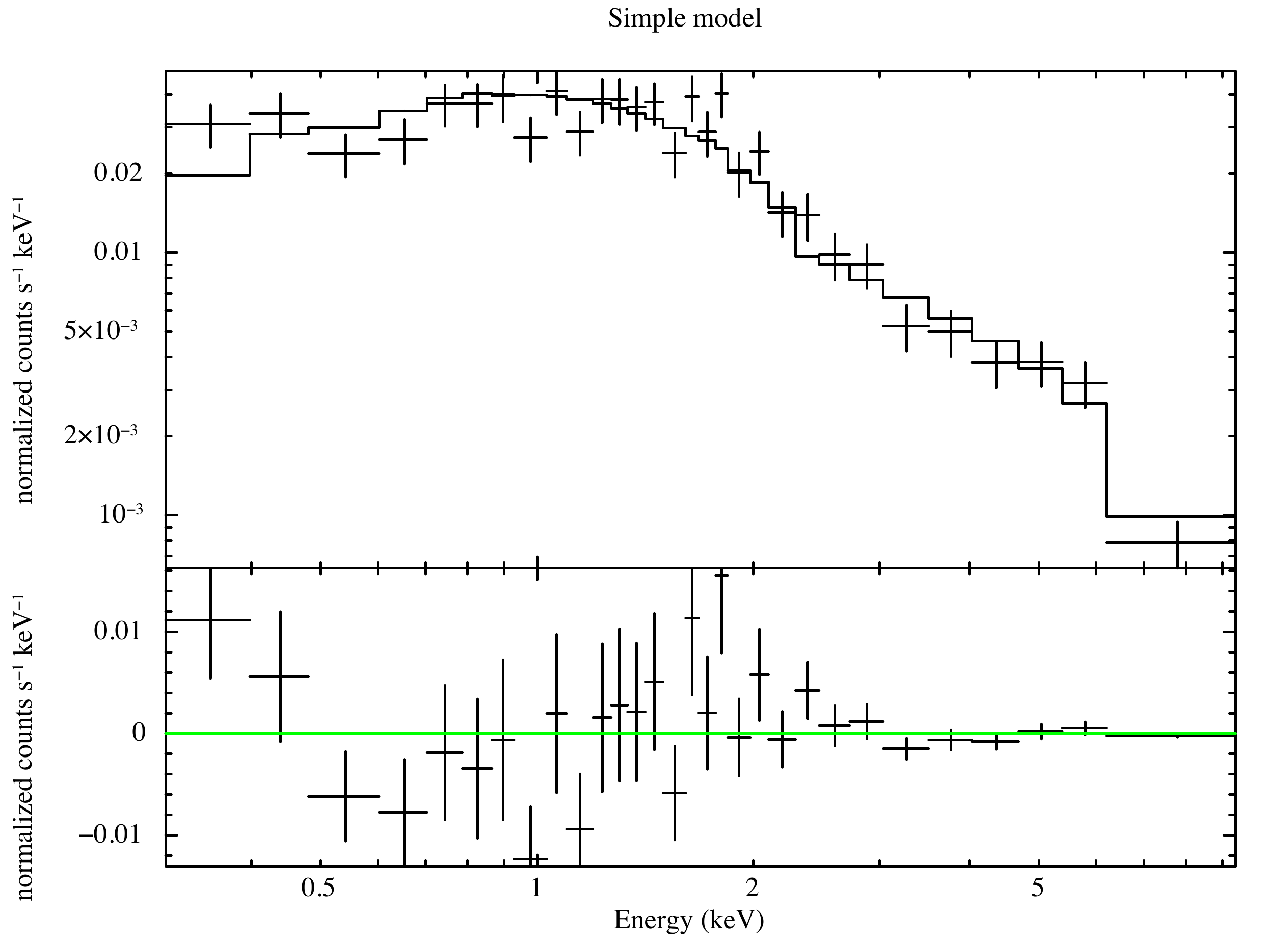}
\includegraphics[width=\columnwidth]{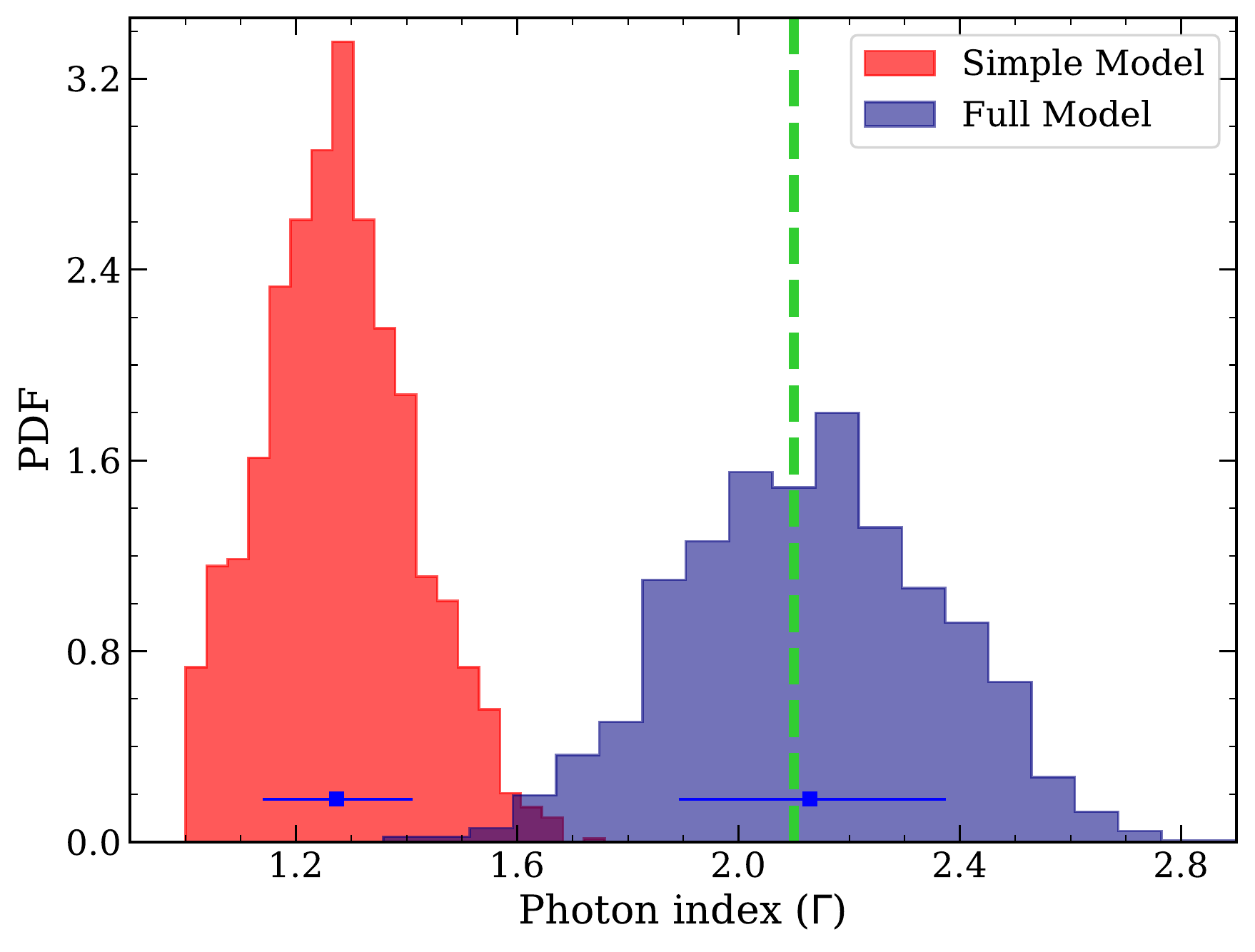}
\includegraphics[width=0.97\columnwidth]{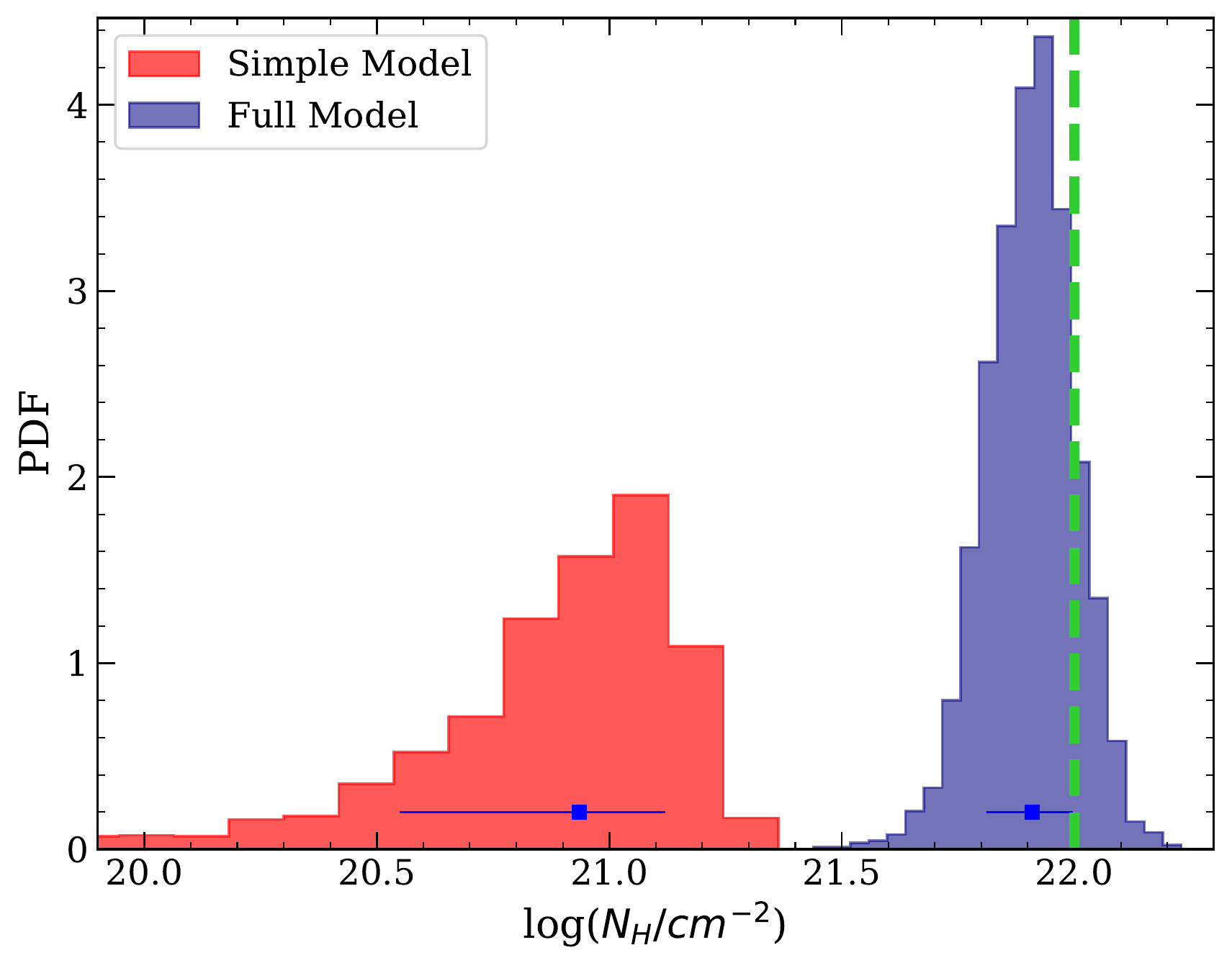}
\caption{Comparison between fits with models of different complexity. Top panels: Best fit of a simulated spectrum using the full model (left) and the simple model (right). The residuals are plotted below. Bottom panels: Posteriors of $\Gamma$ (left) and  $N_\mathrm{H}$ (right) fitted with the full model (light purple) and the simple model (red). The dashed green lines show the input values. The blue points are medians of the posteriors with 68\% credible intervals.}
\label{fig:simplevsfull}
\end{figure*}

To illustrate the effect of hidden parameters, we simulated a spectrum with \emph{XMM-Newton} based on our full model (Sect.~\ref{full model}) using the \texttt{xspec} command \texttt{fakeit}. The input parameters were $\Gamma = 2.1$, $F=1.0\times 10^{-12}$\,ergs\,cm$^{-2}$\,s$^{-1}$, $N_\mathrm{H}=1.0 \times 10^{22}$\,cm$^{-2}$, $q=0.1$, \textit{kT}~$=0.9$\,keV, $\textit{lumin}=10^{39.5}$\,erg\,s$^{-1}$, $R=0.1$, and $f_\mathrm{scatter}=0.08$. We used the response and auxiliary files of a source in the XXL-1000-AGN sample \citep[3XLSS J022915.6-044217;][]{Fotopoulou2016} as a template, with 10\,ks exposure time and $8.2 \times 10^{-2} \pm 2.9 \times 10^{-3}$\,s$^{-1}$ net count rate. The background was neglected in this case. The spectrum was fit with the full model and a simple model using \texttt{xspec}, as shown in the top two panels of Fig.~\ref{fig:simplevsfull}. The simple model was a power law with photoelectric absorption (\texttt{zphabs} $\times$ \texttt{zcutoffpl}) and had only three free parameters: $N_\mathrm{H}$, $\Gamma$ , and the normalization of the power law. The redshift was fixed to the chosen value (0.1) for both models. The minimum C-stat values of the two fits are close to one another. They are 138.7 for the full model and 152.0 for the simple one, using the same 148 data bins and hence 140 and 145 degrees of freedom (d.o.f), respectively.

We used the Bayesian information criterion \citep[BIC;][]{Schwarz1978}, as well as the Akaike information criterion \citep[AIC;][]{Akaike1974}, to compare these two models. The BIC values were calculated by \( \mathrm{BIC} = k\ln{(n)}-2\ln{(\hat{L})}\), where $k$ is the number of free parameters, $n$ is the number of data points, and $\hat{L}$ is the maximum likelihood. As defined in \citet{Cash1979}, \( \texttt{C-stat} = -2\ln{\hat{L}}\). We can find that the BIC of the full model is 178.7, while that of the simple model is 167.0, which means that the simple model is formally preferred. The AIC gives 154.7 and 158.0, respectively, as calculated by \(\mathrm{AIC=2k-2\ln{\hat{L}}}\), which are not statistically different. We might conclude that fitting with the simple model is adequate based on these results. 

However, in this case, but also in reality, we know that the real model does contain secondary components such as soft excess and reflection. The bottom panels of Fig.~\ref{fig:simplevsfull} show marginalized posteriors of $\Gamma$ and $N_\mathrm{H}$ obtained using MultiNest (see Sect.~\ref{posterior sampling}) with the two models. Noninformative priors were applied for the two main parameters, and the priors of other parameters for the full model are presented in Sect.~\ref{model priors}. The simple model fails to recover the true parameters, fitting an absorbed soft spectrum with an unabsorbed hard solution. The full model, on the other hand, is able to retrieve the correct parameters. In the Bayesian framework, applying the simple model can be interpreted as applying the full model with very strong but incorrect priors that restrict secondary components to be zero. These strong assumptions bias the measurements and induce unrealistically small error bars on the parameters of interest. We conclude that it is necessary to analyze every AGN spectrum with the full model, even when the spectrum does not show clear evidence of the presence of additional components.

\section{Spectral analysis}
\label{bayesian method}

To be able to treat AGN spectra with their full complexity, we developed an alternative fitting approach based on Bayesian statistics. Because a Bayesian analysis can be very computationally expensive, we adapted the whole spectral analysis process, which we describe here in detail. In an analysis of real data, we assume that clean event lists are already available.

\subsection{Spectral extraction}
\label{spectral extraction}

The standard way of extracting AGN X-ray spectra is to directly extract from the event files. Redistribution matrix file (RMF) and ancillary response file (ARF) are computed individually for each source to take the vignetting and detector nonuniformities into account. However, this approach leads to a very complex likelihood function because it requires the extraction and separate fitting of several spectra of the same source, which are taken in different exposures with different instruments on board \emph{XMM-Newton}. For efficiency reasons, we simplified the process by stacking images of all the pointings for a given instrument to make mosaics in different energy bands, depending on the spectral resolution we want. The energy bands we used are listed in Table~\ref{table:10bins}. They can be adjusted. We selected this binning because it is close to a uniform distribution in log-scale, except for two bins, which were intentionally set to 1.3--1.9\,keV and 7.2--9.2\,keV because of the Al\,K and Si\,K fluorescence emission lines at 1.5 and 1.7\,keV for both pn and MOS, and those of Cu\,K$_{\alpha}$, Ni\,K$_{\alpha}$, and Zn\,K$_{\alpha}$ around 8\,keV for pn (see the \emph{XMM-Newton} User Handbook \footnote{\url{https://xmm-tools.cosmos.esa.int/external/xmm_user_support/documentation/uhb/epicintbkgd.html}}). 

Spectral extraction was then performed directly on the mosaics by applying aperture photometry on the narrow-band images listed in Table~\ref{table:10bins}. The MOS1 and MOS2 instruments are considered to be identical, which means that we worked with one pn mosaic and one MOS mosaic. The source spectra were extracted by counting the number of photons within a circular region around each source, and the background spectra were extracted from an annulus located outside of the source region. The neighboring catalog sources were masked to avoid contaminating the background region. The vignetting-corrected exposure maps were used to determine the local effective exposure time, which allowed us to use on-axis RMF and ARF files for all sources. The on-axis response files for pn and MOS were rebinned to match the energy band definition listed in Table~\ref{table:10bins} and then multiplied in each band by the local effective exposure time to take the dependence of the vignetting curve on energy into account. The C++ code implementing our spectral extraction technique is entitled \texttt{xphot,} and it is publicly distributed alongside this series of papers\footnote{\url{https://github.com/domeckert/xphot}}.

The first advantage of this approach is that it is much more efficient to create only one consistent set of RMF and ARF than to create many local ones, especially when multiple pointings are combined. Second, it gives us a consistent energy binning for all the spectra, which, together with the constant RMF and ARF, enables us to perform the spectral fitting without \texttt{xspec} by constructing a single redistribution matrix. This accelerates the fitting process by a large factor. Details are presented in Sect.~\ref{posterior sampling}. The use of this technique comes at the price of losing some information because of the relatively coarse spectral binning. However, the loss of information is relevant only for the brightest sources because in faint sources, the uncertainties on the parameters become large and will hide any difference. Moreover, we show in Sect.~\ref{ngc788} that the impact is minimum even for bright sources. In the typical situation of a survey, most sources have a relatively low S/N, such that the loss of information is negligible.

\begin{table}
\caption{Energy bands used for imaging and spectral extraction. Column (1): Number of the energy bands. Column (2): Lower energy boundary. Column (3) Higher energy boundary.} 
\label{table:10bins}
\centering                        
\begin{tabular}{c c c}      
\hline\hline  
\rule{0pt}{1.2em}\noindent
Band number & Energy\_low & Energy\_high \\
\rule{0pt}{1.2em}\noindent
 & (keV) & (keV) \\
\hline 
\rule{0pt}{1.2em}\noindent
  \phantom{0}1 & 0.4 & \phantom{0}0.6 \\      
  \phantom{0}2 & 0.6 & \phantom{0}0.9 \\
  \phantom{0}3 & 0.9 & \phantom{0}1.3 \\
  \phantom{0}4 & 1.3 & \phantom{0}1.9 \\
  \phantom{0}5 & 1.9 & \phantom{0}2.8 \\
  \phantom{0}6 & 2.8 & \phantom{0}4.2 \\
  \phantom{0}7 & 4.2 & \phantom{0}6.3 \\
  \phantom{0}8 & 6.3 & \phantom{0}7.2 \\
  \phantom{0}9 & 7.2 & \phantom{0}9.2 \\
  10 & 9.2 & 12.0 \\
\hline\hline                    
\end{tabular}
\end{table}

\subsection{Background modeling}
\label{background model}

An important feature of X-ray spectra is that spectral bins typically contain a small number of counts and therefore follow Poisson statistics. To take the properties of the background into account in the fits, most previous works used background spectra obtained from source-free regions and subtracted the rescaled background spectra from the source spectra. However, the difference between two Poisson variables is no longer distributed like a Poisson variable, which can lead to severe biases. Instead of subtracting the background, we describe the background spectrum using a physically motivated model and fit the total spectra of the source and of the background simultaneously using both the source and background components in the former case and only the background components in the latter, preserving the Poisson statistics. We separately model the X-ray background (XB) and the non-X-ray background (NXB), and our background model is described in the following. 

\subsubsection{X-ray background}
\label{xb model}

The XB was generated by photons from X-ray sources other than the source of interest. We modeled it with three components that represent the Local Bubble, the Galactic halo, and the cosmic X-ray background (CXB) as follows (in \texttt{xspec} notation):

\(\texttt{constant} \times ( \texttt{apec} + \texttt{TBabs} \times (\texttt{apec} + \texttt{constant} \times \texttt{powerlaw})).\)

The first constant is a scaling factor that takes the different extraction areas for the source and background spectra into account. It was fixed for each source. The photoelectric absorption caused by our Galaxy was modeled using \texttt{TBabs}, as in the source model (see Sect.~\ref{full model}). The first \texttt{apec} model was the emission from the Local Bubble; its temperature was fixed to 0.11\,keV \citep{McCammon2002} and, being local, was unabsorbed. The other two components were absorbed by the same galactic $N_\mathrm{H}$ column density as the source. The emission from the Galactic halo was taken into account by an absorbed \texttt{apec} with the temperature fixed to 0.22\,keV \citep{McCammon2002}. The absorbed power law represents the cosmic X-ray background, which is thought to result from the unresolved AGN population \citep{Moretti2003,Lehmer2012}. The photon index of this power law was fixed to 1.46 following \citet{Molendi2004}. While the shapes of these three components were fixed, their normalization parameters can be determined for each source using ROSAT data. We extracted the ROSAT spectrum in the survey field, using the HEASARC X-ray background tool\footnote{\url{https://heasarc.gsfc.nasa.gov/cgi-bin/Tools/xraybg/xraybg.pl}} \citep{Snowden2019}, and fit it with the model above. The normalization parameters of the Local Bubble and Galactic halo were then fixed to the best-fit values. The CXB can vary on small scales, however, so that its normalization (the second constant) remained free in subsequent fits to the source spectrum, but was constrained to be close to the best-fit value (see Sect.~\ref{model priors}), which is the only free parameter of the XB model.

\subsubsection{Non-X-ray background}
\label{nxb model}

The NXB consists primarily of secondary electrons caused by the interaction of charged high-energy particles with the spacecraft. Because the high-energy particles are going through the structure, the spatial distribution of this component is almost uniform. We used filter-wheel-closed (FWC) EPIC data to create model images of the NXB. We used the corners of the EPIC detectors, which are located outside of the telescope field of view, to estimate the appropriate particle background levels of each observation. To model the time variation of the NXB, the model NXB images were rescaled to match the count rates measured in the unexposed corners of each pointing. All the resulting images were then stacked to create mosaics from which NXB spectra were extracted in both the on-source and off-source regions. The spectral shape of the NXB was fixed to the prediction of the aforementioned model. Its normalization was left free during the fitting procedure, however, to account for possible variations, in particular those induced by residual soft protons. In addition, bright fluorescence lines (Al and Si at 1.5 and 1.7\,keV in pn and MOS; Cu, Ni, and Zn around 8\,keV in pn) are known to be time variable and hard to model. The corresponding energy bins therefore contain additional variable components, and the NXB in the corresponding energy ranges was allowed to vary independently of the NXB continuum. We parameterized the variations by adding one multiplicative parameter $f_\mathrm{nxblines}$ for each of the three lines (Al, Si, and Cu--Ni--Zn).

\subsection{Likelihood}
\label{likelihood}

We took the selection effects on the spectral analysis into account by incorporating the selection function into the likelihood function. Following \citet{Kelly2007}, we introduced an indicator $I$ that denotes whether a source is detected or not. $I=1$ when the detection is successful, otherwise $I=0$. We assumed here that the selection only depends on the observed count rate ($CR$) in the detection band, so that the selection function can be expressed as $P(I=1|CR)$. Different or more complex selection functions can be applied in a similar way \citep[see Sect. 5 of][]{Kelly2007}. We assumed that the shape of the selection function is an error function parameterized as follows:

\begin{equation}
\label{eq:selfuc}
P(I=1|CR) = 0.5 \times \mathrm{erf} (a \times \log(CR)+b) + 0.5
,\end{equation}

\noindent where $a$ and $b$ are parameters that need to be fitted in different cases (see an example in Sect.~\ref{simulated sample}), and erf is the error function. When this selection function is taken into account, the complete likelihood function can be written as

\begin{equation}
L = P(I=1|\mathrm{CR})P(CR|y)P(y|\Theta)
,\end{equation}

\noindent in which $P(CR|y)$ can be neglected because $P(CR|y)=1$. Considering that we have Poisson-distributed data, $P(y|\Theta)$ can be expressed as

\begin{equation}
P(y|\Theta) = \prod_{i=1}^N (tm_i)^{S_i}e^{-tm_i}/S_i!,
\end{equation}

\noindent where $S_i$ and $m_i$ are the observed count and the predicted count rate of each data point, while $t$ is the exposure time. Similarly to the \texttt{xspec} C-stat statistics, the final log-likelihood can be expressed as
\begin{equation}
\ln L = \ln{P(I=1|CR)} + \sum_{i=1}^N -(tm_i) + S_i -S_i(\ln(S_i)-\ln(tm_i))
.\end{equation}

For each source, we computed the joint likelihood of the pn source+background, pn background, MOS source+background, and MOS background spectra. The selection effect must be included to prevent the model from converging to a source that would not be detected. This can happen when sources are selected in a given energy range, but fitted over a broader energy range, for example, when source detection is carried out in the 2--7\,keV band and the spectra are fitted in the entire 0.5--10\,keV band. Without this complete likelihood, the fit could prefer a solution that works better in the 0.5--10\,keV band, even if it results in an undetectable solution in the 2--7\,keV band. The complete likelihood effectively removes these spurious solutions.

\subsection{Bayesian approach}

\subsubsection{Model priors}
\label{model priors}

\begin{table*}
\centering 
\caption{Model priors. Column (1): Parameter name. Column (2): Types of distributions included in the prior. Column (3) and (4): Minimum and maximum of the parameter range. Column (5)$\sim$(7): Mean, standard deviation, and skewness of the Gaussian or skewed Gaussian distribution, if presented in the prior. Column (8): Unit of the parameter. $f_\mathrm{nxblines}$ is used for each emission line in the background model (three in total).}
\label{table:priors} 
\begin{tabular}{c c c c c c c c}        
\hline\hline
\rule{0pt}{1.2em}\noindent
Name  & Type & Min & Max & $\mu$ & $\sigma$& $\alpha$ & Unit \\     
\hline
\rule{0pt}{1.2em}\noindent
  $\Gamma$ &  Uniform + 2 half Gaussian & 1.0 & 3.0 & 1.5, 2.5 & 0.1, 0.1 && \\
  $N_\mathrm{H}$ & Jeffreys & $5 \times 10^{19}$ & $5 \times 10^{24}$ &  &  & &cm$^{-2}$\\    
  $q$ &  Uniform + Skewed Gaussian & 0 & 1.8 & 0.16 & 0.5 & 5.0& \\
  \textit{kT} & Gaussian & 0.01 & 1.2 & 0.5  & 0.4 & & keV\\
  \textit{lumin} & Jeffreys & $10^{38}$ & $10^{41}$ & & &&$\mathrm{erg}$ $\mathrm{s}^{-1}$ \\
  $R$ &  Uniform + half Gaussian & 0 & 2.5 & 1.5 & 0.2 &&\\
  $f_{\mathrm{scatter}}$ &Jeffreys & 0.01\% & 10\% &  &  &&\\
\hline
\rule{0pt}{1.2em}\noindent
  $f_{\mathrm{xb}}$ & Jeffreys & 0.4 & 2.5 &  &  && \\
  $f_{\mathrm{pnnxb}}$ & Gaussian & 0.85 & 1.50 & 1.04 & $5.67 \times 10^{-2}$ &&\\
  $f_{\mathrm{mosnxb}}$ & Skewed Gaussian & 0.9 & 2.5 & 1.08 & 0.27 & 9.14&\\
  $f_\mathrm{nxblines\_1, 2, 3}$ & Jeffreys & 0.1 & 10 &  &  & &\\
\hline\hline                                  
\end{tabular}
\end{table*}

\begin{figure*}[h!]
\centering
\includegraphics[width=17cm]{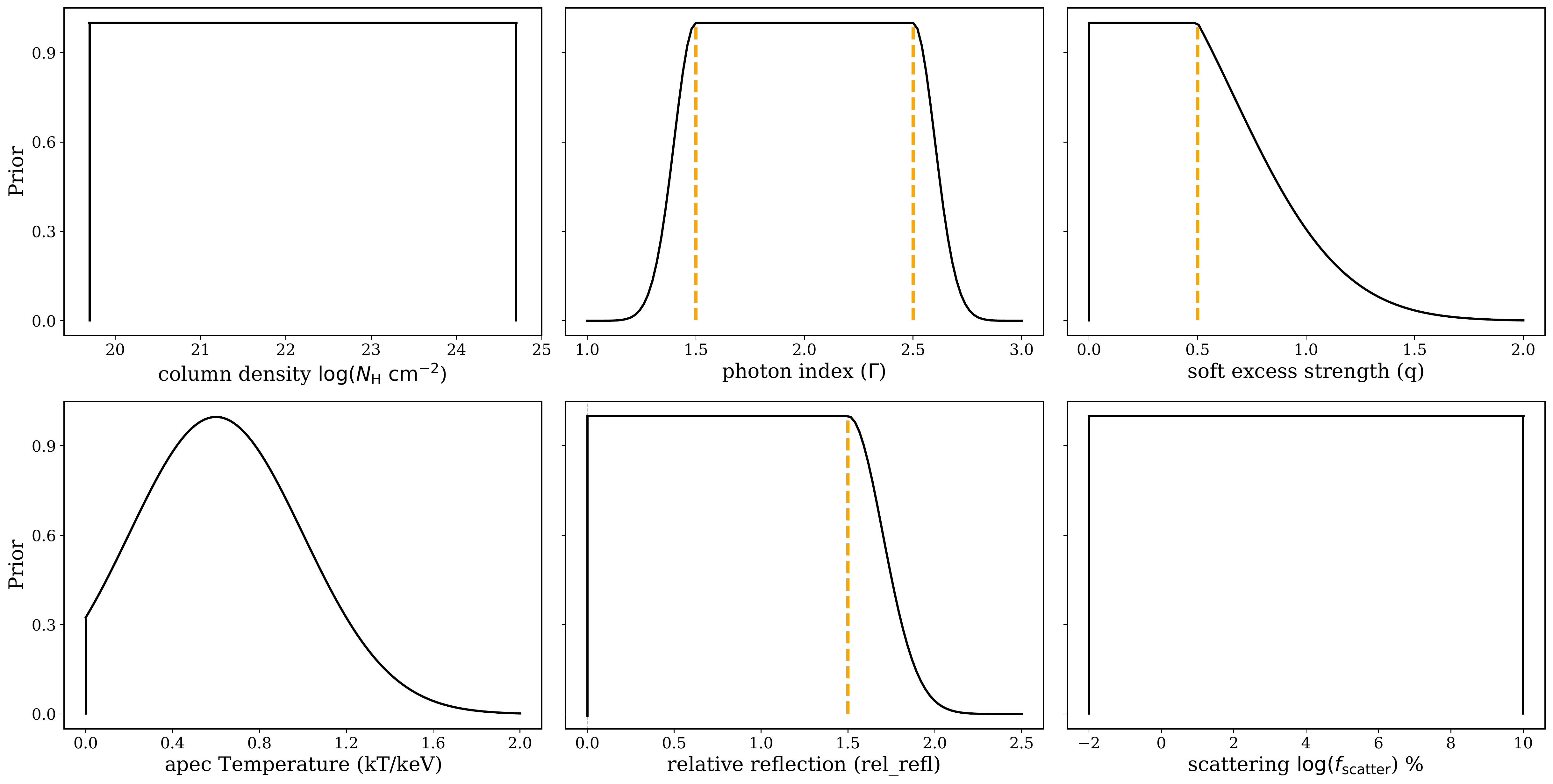}
\caption{Priors of $N_\mathrm{H}$, $\Gamma$, $q$, \textit{kT}, $R,$ and $f_\mathrm{scatter}$. The dashed orange lines point out the locations where the priors switch to different analytical expressions. See Table~\ref{table:priors} for details.}
\label{fig:source priors}
\end{figure*}

The parameter set $\Theta$ is composed of a number of parameters $\theta_i$. If all parameters are independent, Eq.(\ref{eq:bayes}) becomes

\begin{equation}
P(\Theta|y) = P(\{\theta_i\}|y) = \frac{\prod_{i} P(\theta_i) \cdot P(y|\Theta)}{P(y)}
\label{eq:bayes_idp}
.\end{equation}

The priors we used are listed in Table~\ref{table:priors}, and those of the source model are plotted in Fig.~\ref{fig:source priors}. We applied noninformative priors for $\Gamma$ and $N_\mathrm{H}$ because they are the main parameters that we wish to determine. The prior on $\Gamma$ is flat between 1.5 and 2.5 with two Gaussian cuts with $\sigma=0.1$ at both ends in order to avoid boundary effects. A Jeffreys prior (flat in log-scale) was used for $N_\mathrm{H}$ between $5\times 10^{19}$ to $5\times 10^{24}$ cm$^{-2}$. Outside of this range, the 0.5 to 10\,keV X-ray spectra are very insensitive to the absorption because the impact of absorption is negligible when $N_\mathrm{H}$ is too low, while the spectra are totally absorbed when $N_\mathrm{H}$ is too high.

Informative priors were used for the parameters of the secondary components that are difficult to measure. The priors of $f_\mathrm{scatter}$, $R$ and \textit{kT} were taken from \citet[][see their figures~21, 30 and 31]{Ricci2017}. In the case of $f_\mathrm{scatter}$ , we set a Jeffreys prior with an upper limit of 10\% and a lower limit of 0.01\%. The prior on $R$ is flat between 0 to 1.5 with a smooth Gaussian cut extending farther to 2.5 \footnote{In this paper, $R$ is the physical property and is always positive, but $-R$ is passed to the \texttt{xspec} model to ignore the direct emission in \texttt{pexmon.}}. The prior on the temperature (\textit{kT}) of the thermal component is written as a Gaussian prior centered at 0.5\,keV with $\sigma=0.4$\,keV, and its intrinsic 2--10\,keV luminosity (\textit{lumin}) is constrained by a Jeffreys prior in the range from $10^{38}$ to $10^{41}$ erg\,s$^{-1}$. The prior on the soft-excess strength $q$ was derived from the distribution of the soft-excess strength from \citet[][see their table.A.2]{Boissay2016}. It was fit with a flat distribution below 0.5 and a skew-normal distribution above, which was parameterized by \( \mathrm{skewnorm}(x;\mu,\sigma,\alpha) = 2 \times \mathcal{N}(x-\mu,\sigma) \times \mathcal{N}(\alpha (x-\mu),\sigma) \), where norm is the normal distribution, $\mu=0.16$, $\sigma=0.5,$ and $\alpha=5.0$. Its upper boundary was set to 1.8. The skew-normal distribution was implemented with the Python package \texttt{scipy} \citep{scipy2020}.

The XB model has just one free parameter, which is the normalization of the CXB, and we constrained it with a Jeffreys prior from 0.4 to 2.5 because it is a relative variation with respect to the average value we measure using ROSAT data (see Sect.~\ref{background model}). The priors on NXB continuum normalization parameters were obtained from an analysis of about 500 archival \emph{XMM-Newton} extragalactic blank-field observations \citep{Ghirardini2018}. The multiplicative factor for the MOS NXB spectra was found to have an average value of 1.2 and a skewed distribution toward a higher value. We describe it as a Gaussian with positive skewness, which is \( \mathrm{skewnorm}(x;\mu,\sigma,\alpha) \) as defined above, where $\mu=1.08$, $\sigma=0.27$ and $\alpha=9.14$. In the case of the pn, the multiplicative factor is symmetric and centered on unity. We therefore describe it as a simple Gaussian, where $\mu=1.04$, $\sigma=0.06$. We used Jeffreys priors for the multiplicative parameters of the fluorescence lines ($f_\mathrm{nxblines}$), which are bounded between 0.1 and 10.

\subsubsection{Posterior sampling}
\label{posterior sampling}

The posterior was computed by applying the priors on the likelihood following Eq.~(\ref{eq:bayes}). We sampled the posterior using \texttt{MultiNest} \citep{Feroz2008,Feroz2009,Feroz2019}, which is an implementation of the nested-sampling algorithm \citep{Skilling2004}. The main advantages of \texttt{MultiNest} over more common sampling approaches such as a Markov chain Monte Carlo (MCMC) method are the capabilities to sample multimodal posteriors and compute the Bayesian evidence for model selection.

Our choice of extracting spectra from mosaics in fixed energy bins allowed us to implement the computation of the posterior much more efficiently, without using \texttt{xspec}. Because all objects have the same ARF and the same RMF (for a given instrument, pn or MOS), we can convolve the models once for all and create a large model grid, from which counts can be predicted using interpolations without additional convolution. We find an improvement in speed for our Python interpretation by a factor of 10 compared to using the BXA Python interface to \texttt{MultiNest} and \texttt{xspec} \citep[Bayesian X-ray Analysis;][]{Buchner2014}.

We tabulate each spectral component used in our full model with the parameters described in Sect.~\ref{source model}. The details of the parameters of the grid are listed in Table~\ref{table:grid}. The model count rates for each set of parameters in the grid were computed using the \texttt{xspec} command \texttt{fakeit} with the proper ARF and RMF.
The whole fitting process is implemented in Python using \texttt{PyMultinest} \citep{Buchner2014}, which provides a Python interface to the \texttt{MultiNest} library.

\begin{table}
\caption{List of intervals of spectral parameters. Column (1): Parameter name. Column (2): Spacing type of the grid. Column (3) and (4): Minimum and maximum of the parameter. Column (5): Number of steps of the grid.}             
\label{table:grid}    
\centering                        
\begin{tabular}{c c c c c}      
\hline\hline 
\rule{0pt}{1.2em}
Name & Method & Min & Max & \# of steps \\   
\hline 
\rule{0pt}{1.2em}
   z & log & $1 \cdot 10^{-3}$ & 4.0 &  70\\      
   $N_{\mathrm{H}}$ source & log &  $5 \cdot 10^{19}$ & $5 \cdot 10^{24}$ & 50 \\
   $\Gamma$ &  linear & 1.0 & 3.0   & 20 \\
   $N_{\mathrm{H}}$ galactic & linear & 0.01 & 0.03    & 20 \\
   kT & linear & 0.01 & 1.20    & 60 \\
   R &  linear & 0 & 2.5  & 2 \\ 
\hline\hline                     
\end{tabular}
\end{table}

\subsubsection{Spectral analysis of NGC\,788}
\label{ngc788}

\begin{figure}[tb]
\centering
\includegraphics[width=\columnwidth]{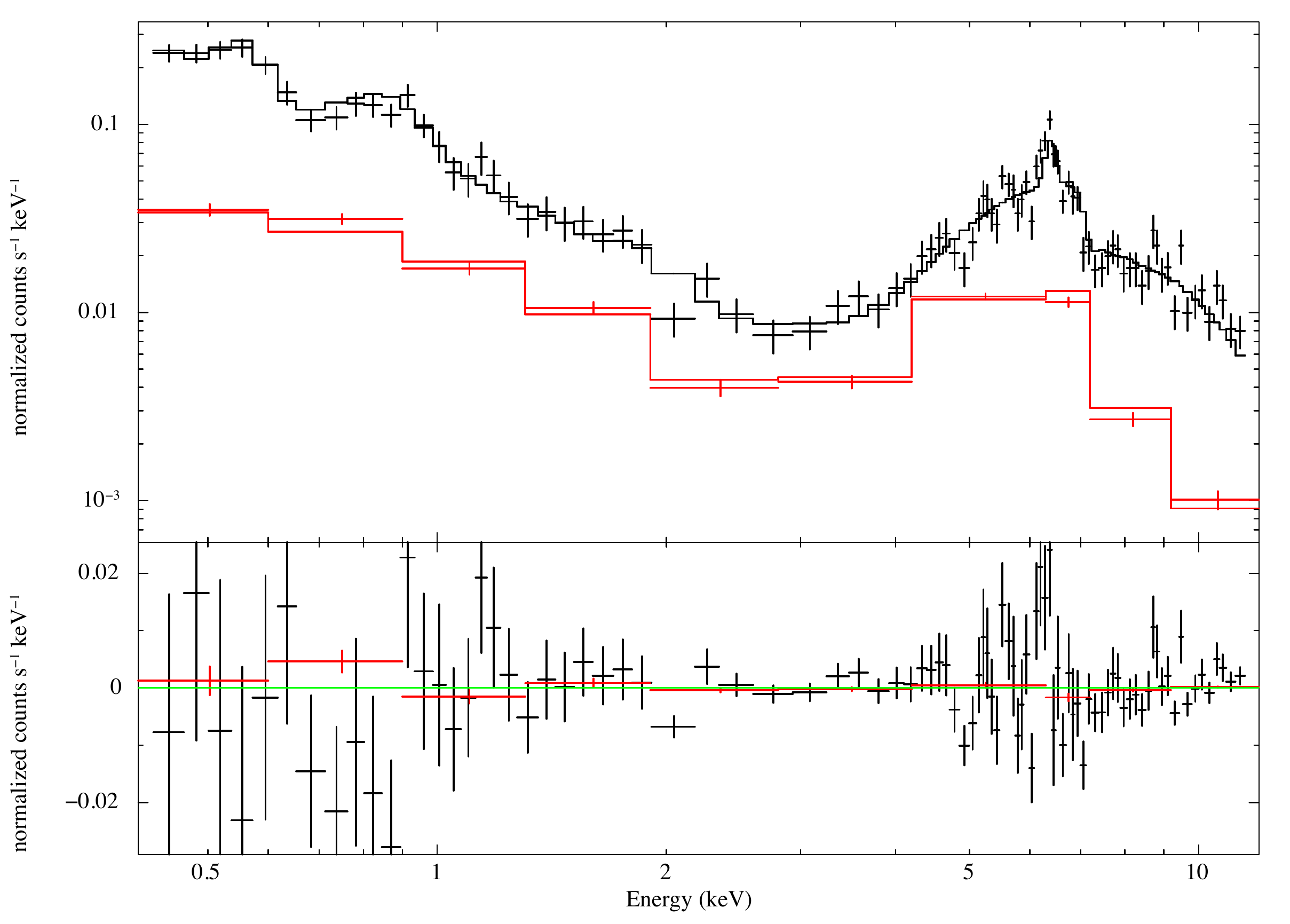}
\caption{Joint fitting of the spectra of NGC\,788 extracted in two ways. The pn spectrum that is directly extracted in a standard way from event files, which has the full resolution, is shown in black, and the MOS spectrum extracted from aperture photometry in ten bins is shown in red. Together with the 10-bin pn spectrum and full-solution MOS spectrum, they are simultaneously fit to our source plus background model.}
\label{fig:jointfit}
\end{figure}

We tested our spectral analysis method on the bright source \object{NGC\,788} (lower panel of Fig.~\ref{fig:realfit}) and compared the results with what was obtained using a standard fit to the full spectrum (see Sect.~\ref{real example}). Raw data (Obs. ID: 0601740201) were downloaded from the \emph{XMM-Newton} Science Archive. Following the data analysis pipeline described in detail in section~2.2 of \citet{Ghirardini2019}, we used \texttt{emchain} and \texttt{epchain} of XMMSAS v13.5 to extract calibrated event files, and then applied the executables \texttt{mos-filter} and \texttt{pn-filter} to filter out time periods affected by soft-proton flares. Vignetting-corrected exposure maps were generated with \texttt{eexpmap} and count images were extracted from the three EPIC detectors in ten energy bins. The on-axis RMF and ARF files were generated with \texttt{rmfgen} and \texttt{arfgen}. The 10-bin source and background spectra of pn and MOS were then extracted from the images as described in Sect.~\ref{spectral extraction}. The radius of the source region is 40 arcsec and the inner and outer radii of the background region are 40 and 70 arcsec. Full-resolution spectra of pn and MOS were extracted following the standard SAS procedure to extract spectra of point-like sources using \texttt{evselect}. The same source and background regions were applied in both analysis.

\begin{figure*}[tb]
\centering
\includegraphics[width=17cm]{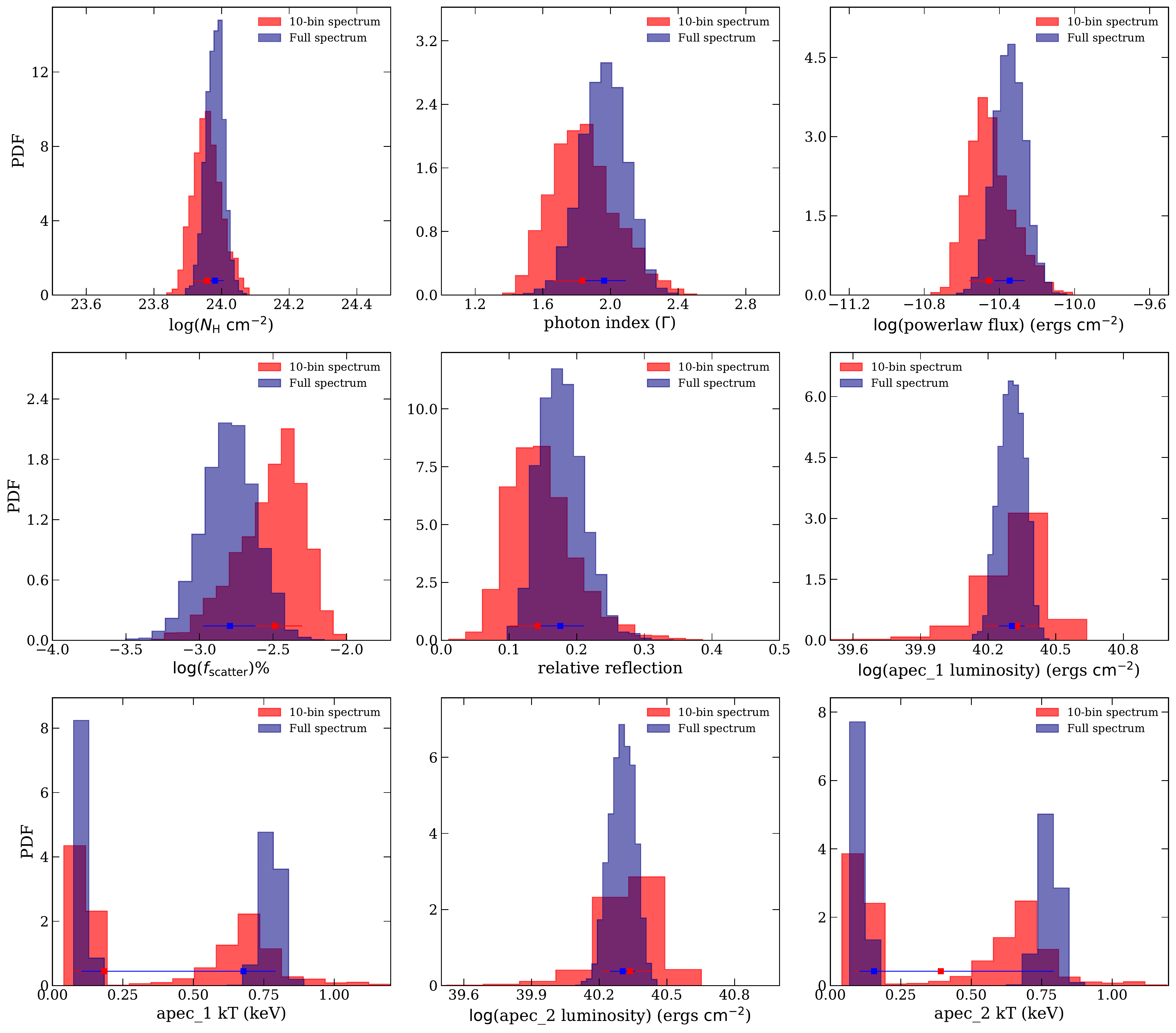}
\caption{Marginalized posteriors of the spectral parameters sampled using the 10-bin spectra (red) and the full spectra (purple). Error bars are the medians of the PDFs with 68\% credible interval.}
\label{fig:pdf_compare}
\end{figure*}

We first fit the full spectra and the 10-bin spectra of both pn and MOS simultaneously in \texttt{xspec} using our full model. We added a second \texttt{apec} in the source model in this case to model residuals that are clearly detected in the spectra because of the high-S/N of this source. We were unable to generate our NXB models for the full-resolution spectra because they require binned images. We therefore modeled them using a power law plus two Gaussians centered on 1.6 and 8.0\,keV, and then fit them to the 10-bin NXB. When we fit the full-resolution source+background spectra, parameters of this model were fixed to the best-fit values, and the same parameters of NXB as described in Sect.~\ref{background model} were added to the fixed model and left free to vary. Fig.~\ref{fig:jointfit} shows the simultaneous fit to the full pn and MOS spectra and to the 10-bin pn and MOS spectra. For clarity, only the full-resolution pn and the 10-bin MOS spectra are shown. The two spectra can be simultaneously fit with the same model and same parameter values, although the high-resolution spectrum shows complex residuals around the Fe K$_{\alpha}$ line. 

To compare the results of the fitted parameters and their uncertainties, we used \texttt{MultiNest} to fit the two sets of spectra separately. For the full spectra, we used BXA to perform the spectral analysis in the \texttt{xspec} environment, while for the 10-bin spectra, we followed our method presented above and precomputed the model grid. The priors used for the parameters were the same in both cases (identical to Sect.~\ref{model priors}). Marginalized posteriors of the two approaches are shown in Fig.~\ref{fig:pdf_compare}. The two methods obtain consistent results. Although the PDFs of the full spectra are more narrowly distributed than those of the 10-bin ones, which is expected because of the small loss of information due to rebinning, our method using the 10-bin spectra is able to obtain similar measurements and even good constraints for parameters of secondary components, such as the reflection and thermal-emission components. \citet{Ricci2017} used a model (B3) very similar to our full model to fit the spectra of NGC\,788. They found in general comparable results with our fitted values, for instance, $\log(N_\mathrm{H})=23.77\pm0.05$, which agrees in general with our results. On the other hand, they find $\Gamma=1.64\pm0.26$, which is slightly lower than our fitted values. However, this difference can be explained by the fact that $\Gamma$ has been determined using a fit that also makes use of Swift/BAT data and a free cutoff energy that is found to be around 70 keV. When the cutoff energy was fixed to 500 keV, \citet{Ricci2017} obtained $\Gamma=1.86\pm0.08$, which is consistent with our results.

This example shows that even at high S/N, the results of our spectral analysis method match those of the standard method, and the loss of information due to the binning of the spectra is minimum. The computation efficiency of our method is also demonstrated by this example: the sampling of the posterior using BXA for the full spectra takes more than 6 hours, but only takes about 20 minutes using our method. It is extremely expensive to use the full model, and it is much more difficult to apply it to very large samples.

\subsection{Maximum-likelihood approach}
\label{ML approach}

\begin{table}
\caption{List of additive models used in the six models with different complexities in the ML approach. Column (1): Additive components. Column (2--7): M1 -- M6.}             
\label{table:ml_models}    
\centering                        
\begin{tabular}{c c c c c c c}      
\hline\hline
\rule{0pt}{1.2em}\noindent
Model component & M1 & M2 & M3 & M4 & M5 & M6 \\   
\hline
\rule{0pt}{1.2em}\noindent
   Power law &  \checkmark & \checkmark & \checkmark & \checkmark & \checkmark & \checkmark\\ 
\noindent\rule{0pt}{1.0em}
   Soft excess & & \checkmark &  & \checkmark & \checkmark & \checkmark\\
\rule{0pt}{1.0em}\noindent
   Reflection & & & \checkmark & \checkmark & \checkmark & \checkmark\\
\rule{0pt}{1.0em}\noindent
   Scattering & & & & & \checkmark & \checkmark\\
\rule{0pt}{1.0em}\noindent
   Thermal emission & & & & & & \checkmark\\
\hline\hline                      
\end{tabular}
\end{table}

For comparison, we present a spectral analysis method using the classical ML approach. In order to maintain high computation efficiency, we used our table models to calculate model spectra. Source plus background and background spectra extracted from the mosaics were fit together in order to preserve the Poisson statistics. The same likelihood function as shown in Sect.~\ref{likelihood} was applied here.

To fit low-S/N data with a ML approach, there are two common practices. The first is to simply apply an absorbed power law for all the sources \citep{Perola2004,Tajer2007}. The second is to start with a simple power law, and then, depending on the quality of the fit, decide to apply a more complex model or not. The more complex model is selected depending on the characteristics of the spectra. However, because there is no direct estimate of the goodness-of-fit (GoF) for C-stat, it is not trivial how to assess when a more complex model is needed. Some authors bin the spectra such that each energy bin has more than 10--30 counts to approximate a $\chi^2$ distribution, for which a GoF exists, and a model selection can be made with an F-test \citep{Mateos2005, Corral2011, Brightman2011}. Others simply rely on a qualitative assessment, such as a visual inspection of the residues and of the deviations from the simple best fit \citep{Tozzi2006, Shinozaki2006}.

We propose here a statistically sound approach that can be applied automatically. We implement a ML approach as follows: we first define six different source models, using the components defined in Sect.~\ref{source model}, with different complexities as M1 to M6 listed in Table~\ref{table:ml_models}. The six models have 27, 26, 26, 25, 24, and 22 free parameters. We fit the spectra with all the models, and find the best-fit parameters for each of them. The optimization is performed using the MINUIT library \citep{James1975} via its \texttt{iminuit} Python interface. The best model for each source is then selected using the corrected Akaike information criterion \citep[AICc;][]{Cavanaugh1997}, which is the modified AIC for small sample sizes and calculated by \( \mathrm{AICc} = \mathrm{AIC} + (2k^2+2k)/(n-k-1) \), where \( \mathrm{AIC} = 2k - 2ln(\hat{L}) \). $\hat{L}$ is the maximum likelihood value of the model; $k$ and $n$ denote the number of parameters and the sample size, which is the number of data points of the spectra, respectively. We then simply select the model with the lowest AICc value.

\section{Simulations}
\label{simulation}

We compared the results of the Bayesian analysis we have described above with those obtained with the ML approach. For this purpose, we simulated a realistic sample of AGN and created representative source and background spectra. Finally we applied both our Bayesian spectral-analysis method and the classical ML spectral-analysis approach on them. We describe below the simulations and the results in detail.

\subsection{Simulated sample}
\label{simulated sample}

\begin{figure}[h!]
\centering
\includegraphics[width=\columnwidth]{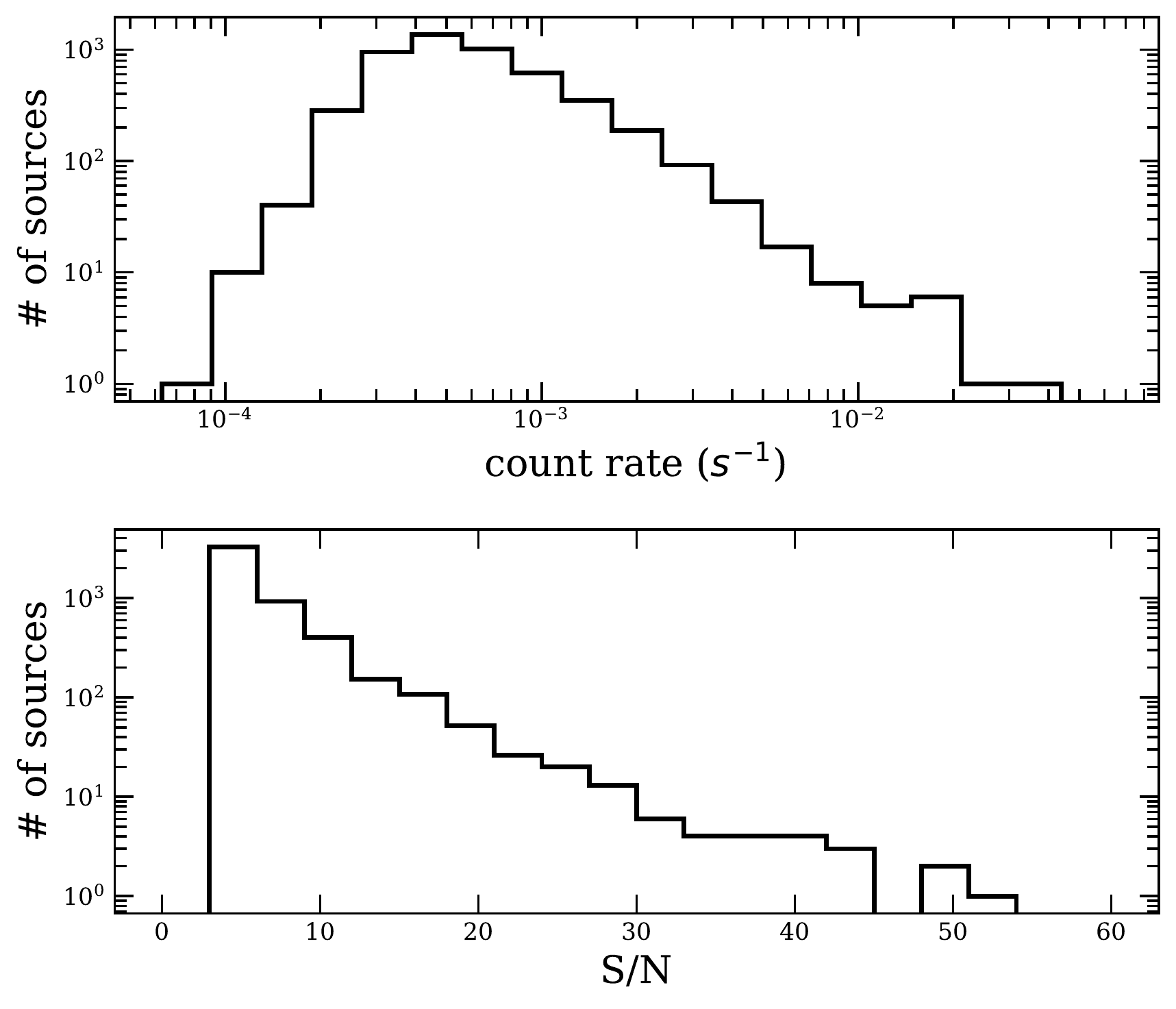}
\caption{Histograms of count rate (top) and S/N (bottom) of the simulated sample in the 2--7\,keV band.}
\label{fig:simul_sample}
\end{figure}

\begin{figure}[tb]
\centering
\includegraphics[width=\columnwidth]{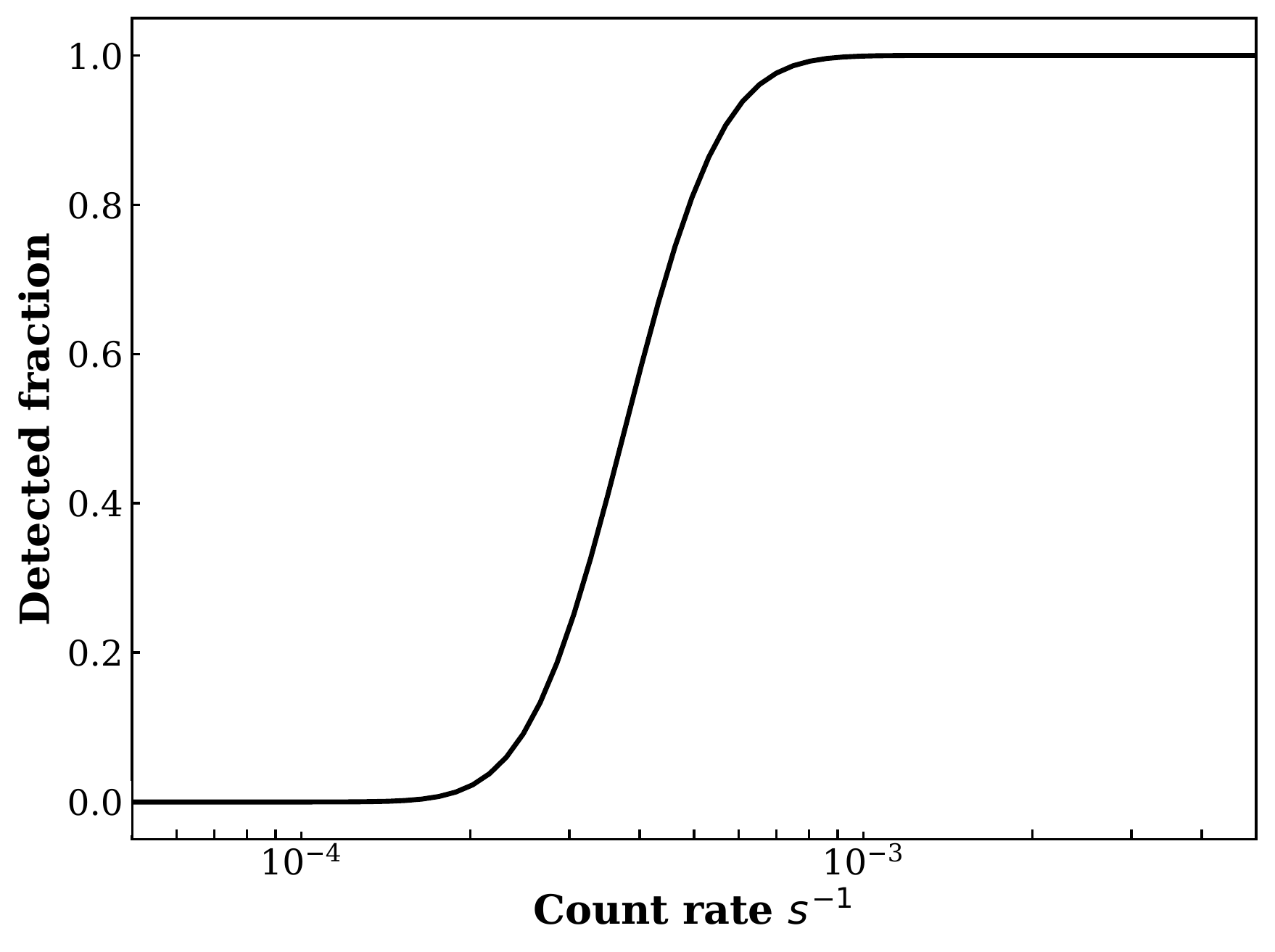}
\caption{Selection function of the simulated sample based on count rate ($\mathrm{s}^{-1}$) .}
\label{fig:selfunc}
\end{figure}

\begin{figure*}[tb]
\centering
\includegraphics[width=17cm]{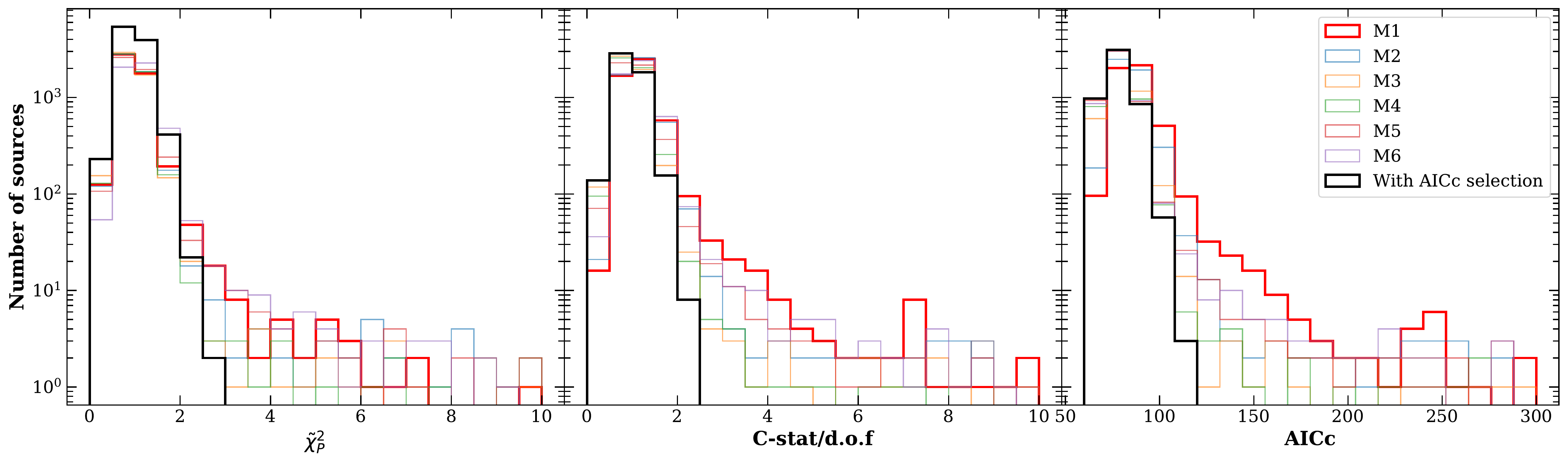}
\caption{Reduced $\chi^2_\mathrm{P}$ (left), C-stat/d.o.f (middle), and AICc (right) distributions of the fits of the simulated sample using ML. The red line shows the results with M1 alone, and the black line shows the results with the AICc selection. The results with M2 -- M5 are also shown with thin lines.}
\label{fig:ml_cstat}
\end{figure*}

We first built a sample of 5000 AGN to a relatively deep depth, and created for each of them pn and MOS spectra with background. The exposure time was randomly drawn from a log-normal distribution $10^t$\,ks, with $t\sim {\cal N}(2, 0.225)$, where 2 is the mean and 0.225 is the variance. Our full model (Sect.~\ref{source model}) was used to simulate the source spectra. $\Gamma$ was drawn from a Gaussian distribution with mean $\mu = 1.95$ and standard deviation $\sigma = 0.15$ \citep{Nandra1994,Buchner2014}. $N_\mathrm{H}$, $F$, $q$, \textit{kT}, \textit{lumin}, $R,$ and $f_\mathrm{scatter}$ were drawn from their priors described in Sect.~\ref{model priors}. The galactic hydrogen column density was fixed at $1.74 \times 10^{20} \mathrm{cm}^{-2}$. The redshift was converted from the comoving distance, which was randomly drawn in the comoving volume up to redshift 4, assuming AGN are uniformly distributed in space. The cosmology we assumed is the WMAP9 $\Lambda$CDM: $H_0=69.3$ km/s/Mpc, $\Omega_\mathrm{\Lambda}=0.72$, $\Omega_\mathrm{m}=0.28$, and $\sigma_\mathrm{8}=0.82$ \citep{Hinshaw2013}. To calculate the intrinsic rest-frame 2--10\,keV luminosity ($L_\mathrm{X}$), we only considered the unabsorbed primary power law and reflection, as the other components contribute mostly below 2\,keV, for example, soft excess, or are much smaller than the primary power law when there is no absorption, for example, scattering. Simulated objects were then selected so that $L_\mathrm{X}$ follows the X-ray luminosity function (LXF) of \citet{Aird2015}. We used the flexible double power-law model parameterized in the hard band.

XB was simulated based on the XB model defined in Sect.~\ref{background model}, and NXB was simulated using a random NXB spectrum of the \emph{XMM}-COSMOS sample described in Paper~\rom{2} as a template. Background parameters were also randomly drawn from their priors. With the source and background spectra and the exposure time, we calculated the photon counts in the 2--7\,keV band as well as the S/N, defined as $(N_\mathrm{s} - N_\mathrm{b}/r)/ \sqrt{N_\mathrm{s} + N_\mathrm{b}/r^2}$, where $N_{\mathrm{s}}$ is the total number of counts in the source region, $N_{\mathrm{b}}$ is that in the background region, and $r$ is the ratio of the background and the source area. We applied a simple selection at S/N$> 3$ to reject undetectable source. The final count rate and S/N distributions of the simulated sample are shown in Fig.~\ref{fig:simul_sample}. We measured the fraction of selected sources after applying the S/N cut as a function of count rate, and fit it using the selection function presented in Sect.~\ref{likelihood}. We obtain the parameters $a = 5.2 \pm 0.9$ and $b = 3.42 \pm 0.02$. The fitted selection function is shown in Fig.~\ref{fig:selfunc}.

Due to the strong selection effect for highly absorbed sources, there are not enough Compton-thick (CT; i.e.,\ sources with $N_\mathrm{H}\ge 1.6 \times 10^{24}$ cm$^{-2}$) sources in the simulated sample above. To better compare the performance of different spectral fitting methods in the CT regime, we simulated another sample of 500 CT AGN by drawing $N_\mathrm{H}$ between $10^{24}$ and $5 \times 10^{24}$ cm$^{-2}$, everything else being unchanged. The final $N_\mathrm{H}$ distribution of the total sample is shown in the upper left panel of Fig.~\ref{fig:simul_stat}.

\subsection{Results}
\label{results}

\begin{figure*}[tb!]
\centering
\includegraphics[width=17cm]{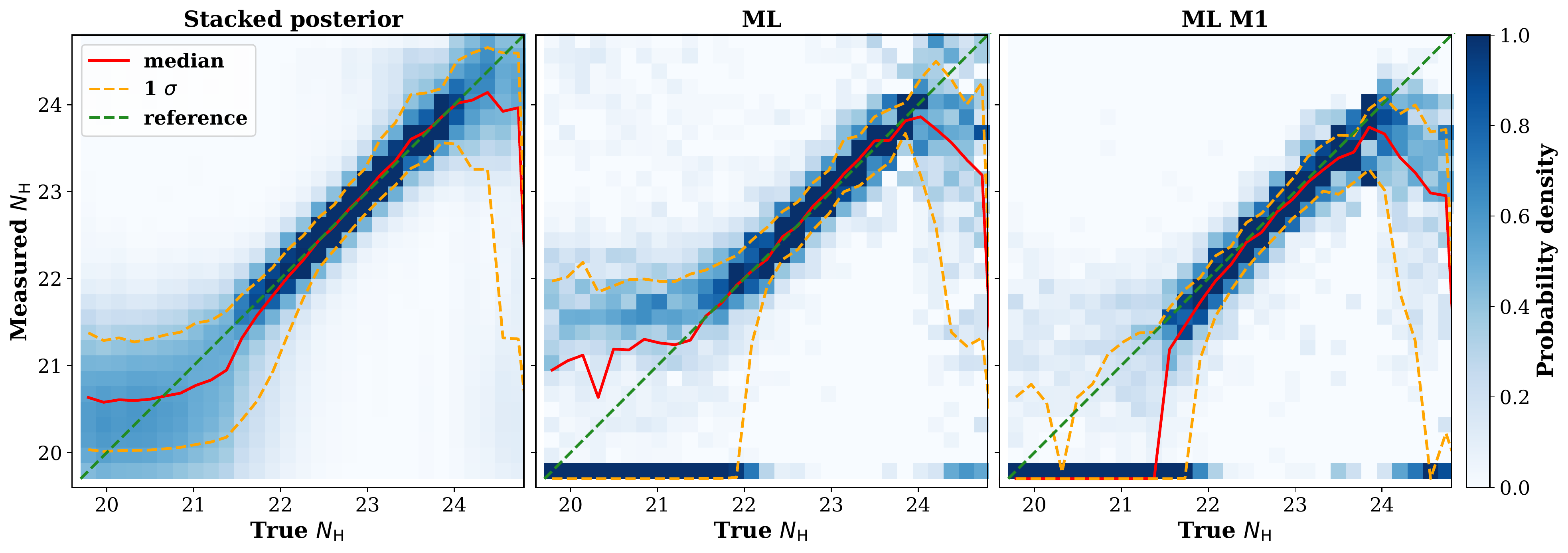}
\includegraphics[width=17cm]{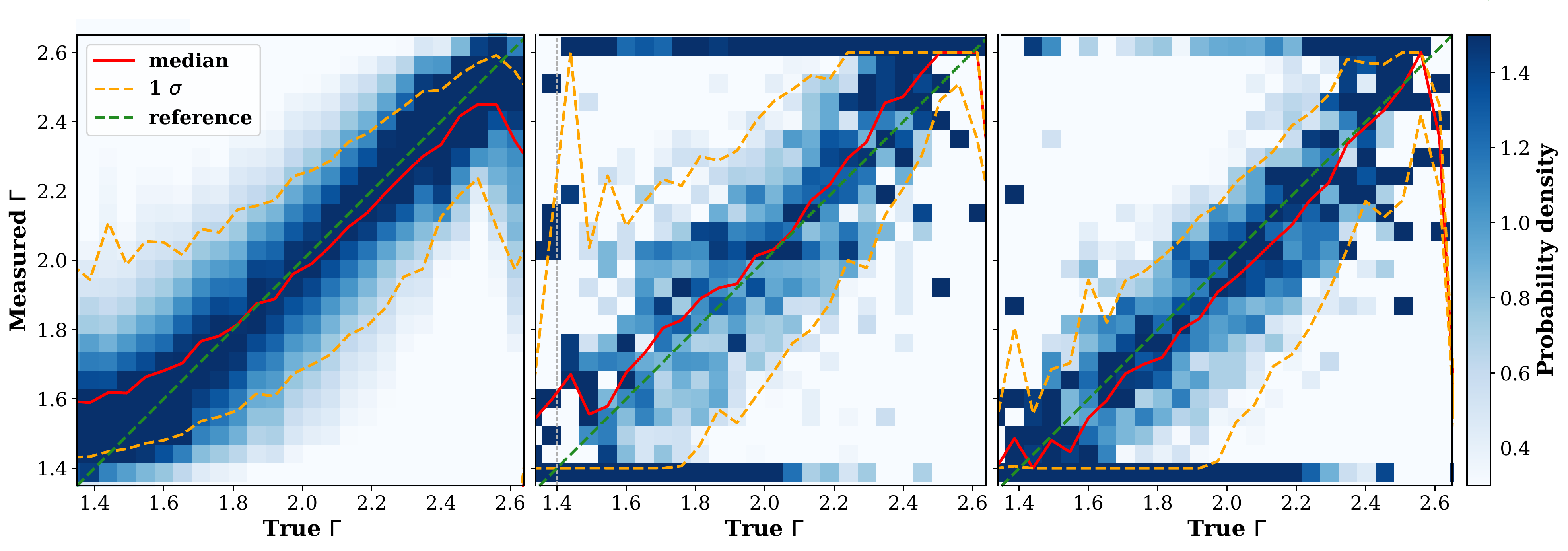}
\includegraphics[width=17cm]{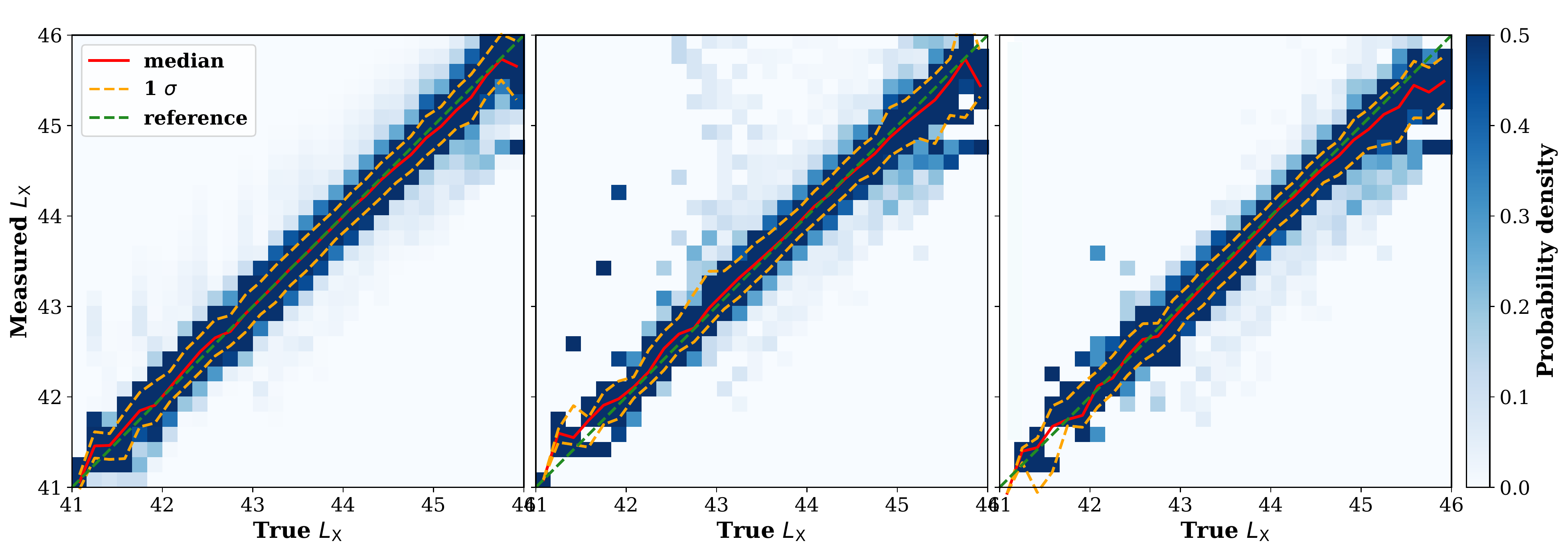}
\caption{Stacked probability density plots of $N_\mathrm{H}$, $\Gamma$, and $L_\mathrm{X}$ obtained with the different fitting approaches. Left column: Bayesian method. Middle: ML method.  Right: M1 model. The true values are binned to 30 bins. The red line shows the medians of the distribution in each bin, and the dashed orange lines are the 16\% and 84\% percentiles. The dashed green line shows the reference where the fitted value equals the true one.}
\label{fig:simul_density}
\end{figure*}

\begin{figure*}[tb!]
\centering
\includegraphics[width=17cm]{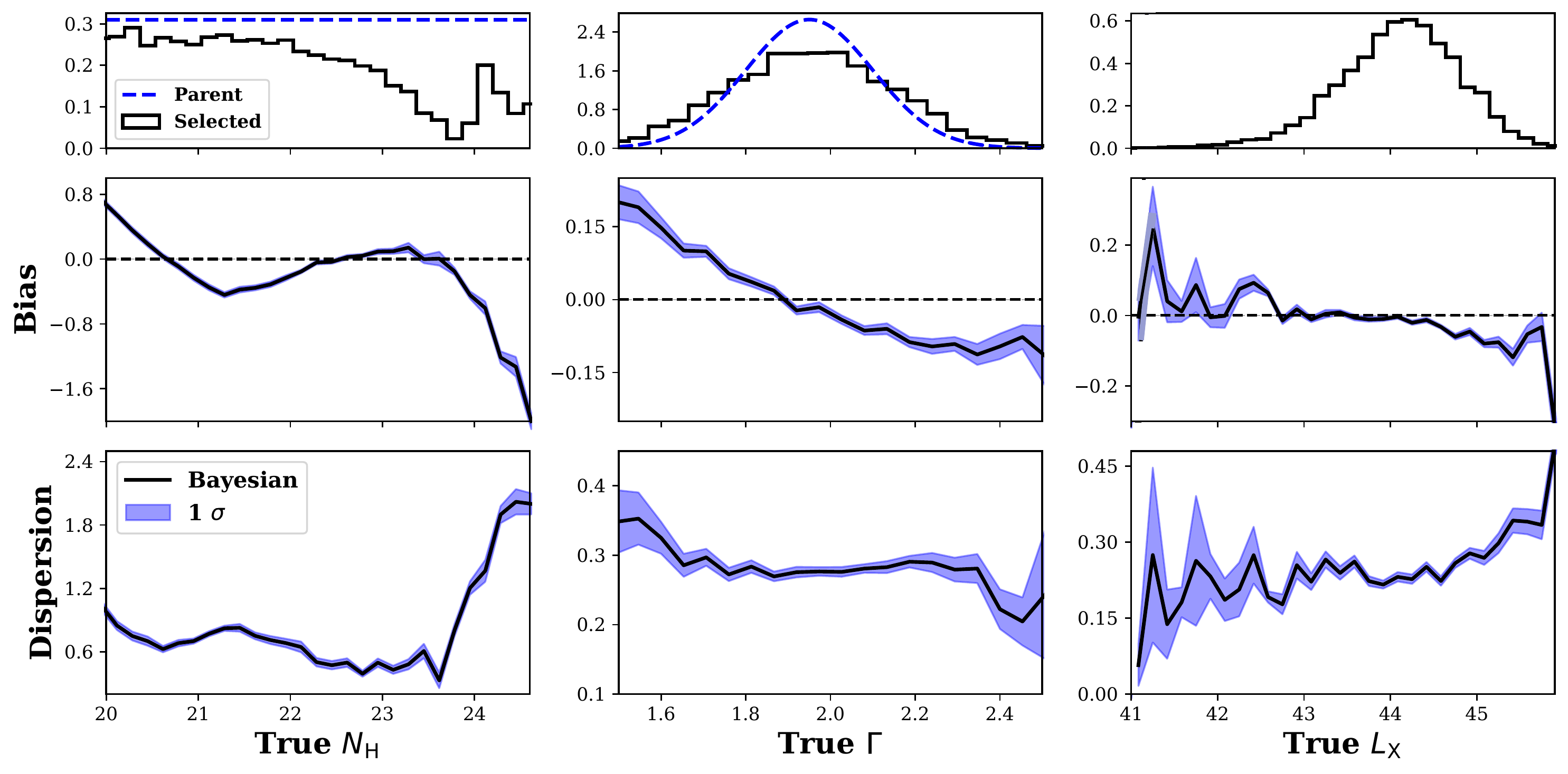}
\caption{Parent distributions of $N_\mathrm{H}$, $\Gamma$, and $L_\mathrm{X}$ (from left to right). Top row: Parent distributions of the main parameters (dashed blue) compared to those after selection (black). The parent distribution of $L_\mathrm{X}$ is not shown because it was computed based on other spectral parameters and then selected by the XLF (see Sect.~\ref{simulated sample}). Middle and bottom rows: Average bias and dispersion with 1$\sigma$ uncertainties (blue area) of the Bayesian method with respect to the true values. In each bin we randomly drew one value from each PDF and then computed the bias and dispersion, and this procedure was repeated for 1000 times to obtain the average bias and dispersion, together with their uncertainties.}
\label{fig:simul_stat}
\end{figure*}

We applied both our Bayesian method and the ML method on the simulated sample. For the ML approach, we find after the model selection that the majority of the sample (3590) prefers M3 (power law plus reflection), while 385, 22, 1022, 472, and 9 sources require M1, M2, M4, M5, and M6 respectively. Figure \ref{fig:ml_cstat} shows the improvement of three estimators after the model selection. The first two are the Pearson-reduced $\tilde{\chi}^2_\mathrm{P}$, which is the $\chi^2$ used for Poisson data with finite counts, and the C-stat divided by the number of d.o.f. These two estimators are not accurate enough for a model selection because they would be valid in the limit of a large number of counts or data points; we can nevertheless qualitatively see the improvement in the fit. All objects with $\tilde{\chi}^2_\mathrm{P}$ or reduced C-stat higher than 3 obtain a better fit after model selection than the results with M1 alone, which shows that more complex models fit these spectra significantly better. The plot in the right panel of Fig.~\ref{fig:ml_cstat} shows the distribution of AICc, which is largely reduced after model selection. This shows that despite the low-to-moderate S/N, the spectra of most objects require not only an absorbed power law, but also a reflection component and sometimes a soft excess as well. On the other hand, models with high complexity (M5 and M6) are rarely required by the data, even though M6 is the correct model. By contrast, the Bayesian method does not require any selection.

We compare the results of the fitting of spectral parameters, especially the primary parameters, for example, $N_\mathrm{H}$, $\Gamma$ and the rest-frame 2--10\,keV intrinsic X-ray luminosity ($L_\mathrm{X}$). We show the results of ML with AICc model selection, as well as M1 alone, because M1 is usually applied to low-S/N data in the literature. The main difference between the ML and Bayesian analysis results is that the ML provides single values for these parameters, while Bayesian analyses provide posterior distributions. For a fair comparison, we therefore computed the two-dimensional histograms of the best-fit ML values versus the true input values for the ML measurements, while we stacked the PDFs for the Bayesian posteriors in each bin. Then we compared the resulting 2D histograms with the 2D stacked PDFs, as shown in Fig.~\ref{fig:simul_density}. 

\begin{figure}[tb]
\centering
\includegraphics[width=\columnwidth]{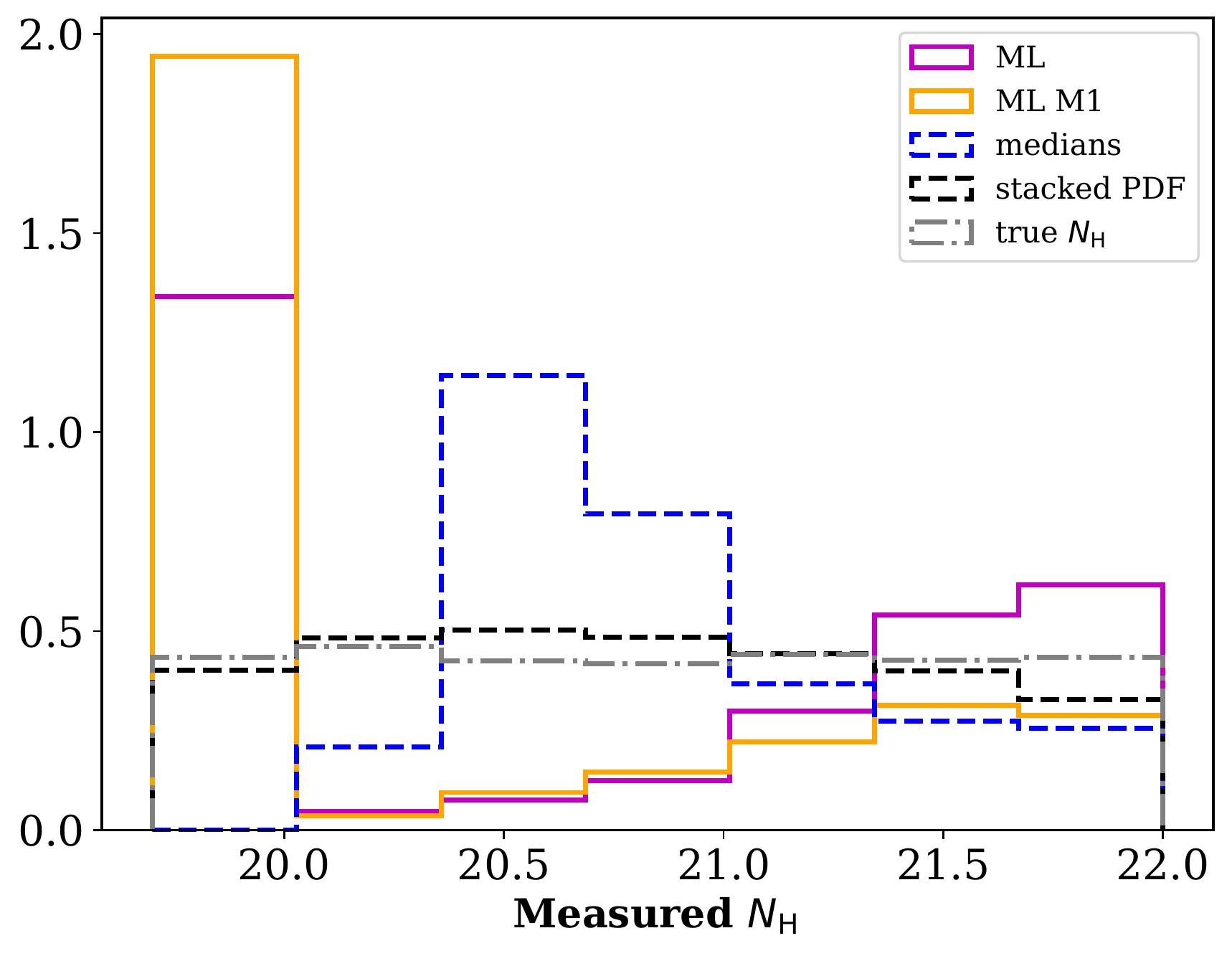}
\caption{Distributions of $N_\mathrm{H}$ in the unabsorbed subsample ($N_\mathrm{H}$\,<\,$10^{22}$\,cm$^{-2}$). The result with the ML approach using AICc selection is shown by the solid magenta line, and the approach using M1 alone is shown by the solid orange line. The results with the Bayesian method using medians and stacked PDF are shown by the dashed blue and black lines, respectively. The gray line shows the parent distribution.}
\label{fig:nh_unabsorbed}
\end{figure}

\begin{figure}[tb]
\centering
\includegraphics[width=\columnwidth]{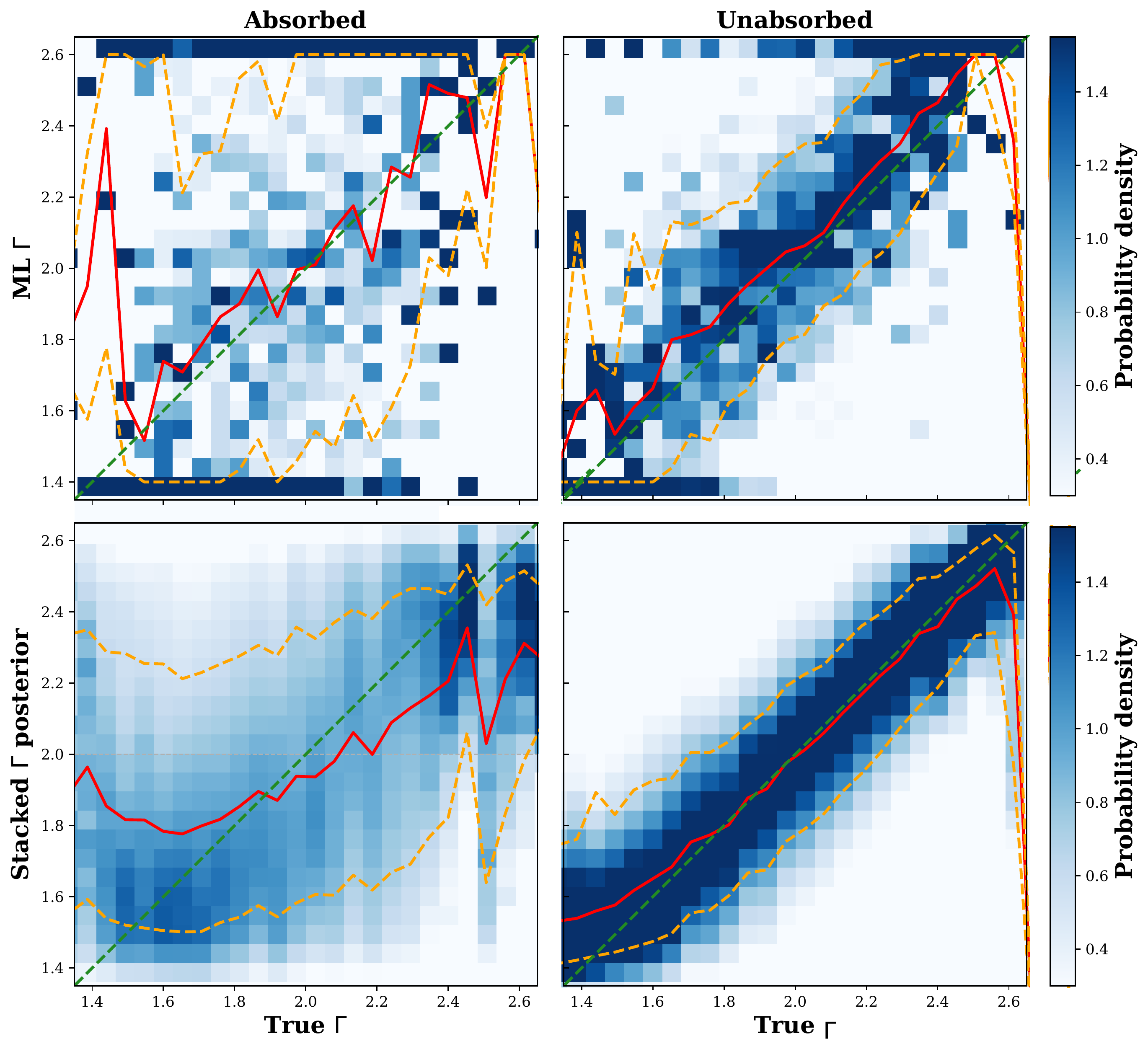}
\caption{Measured $\Gamma$ with ML with AICc model selection (top row) and stacked posteriors (bottom row) against the true value for different subsamples: $N_\mathrm{H}>10^{22} \mathrm{cm}^2$ (absorbed, left column), $N_\mathrm{H}\leq10^{22} \mathrm{cm}^2$ (unabsorbed, right column). The red line shows the medians of the distribution in each bin, and the dashed orange lines are the 16\% and 84\% percentiles. The dashed green line shows the reference where the fitted value equals the true one.}.
\label{fig:simul_density_nh}
\end{figure}

\begin{figure}[tb!]
\centering
\includegraphics[width=\columnwidth]{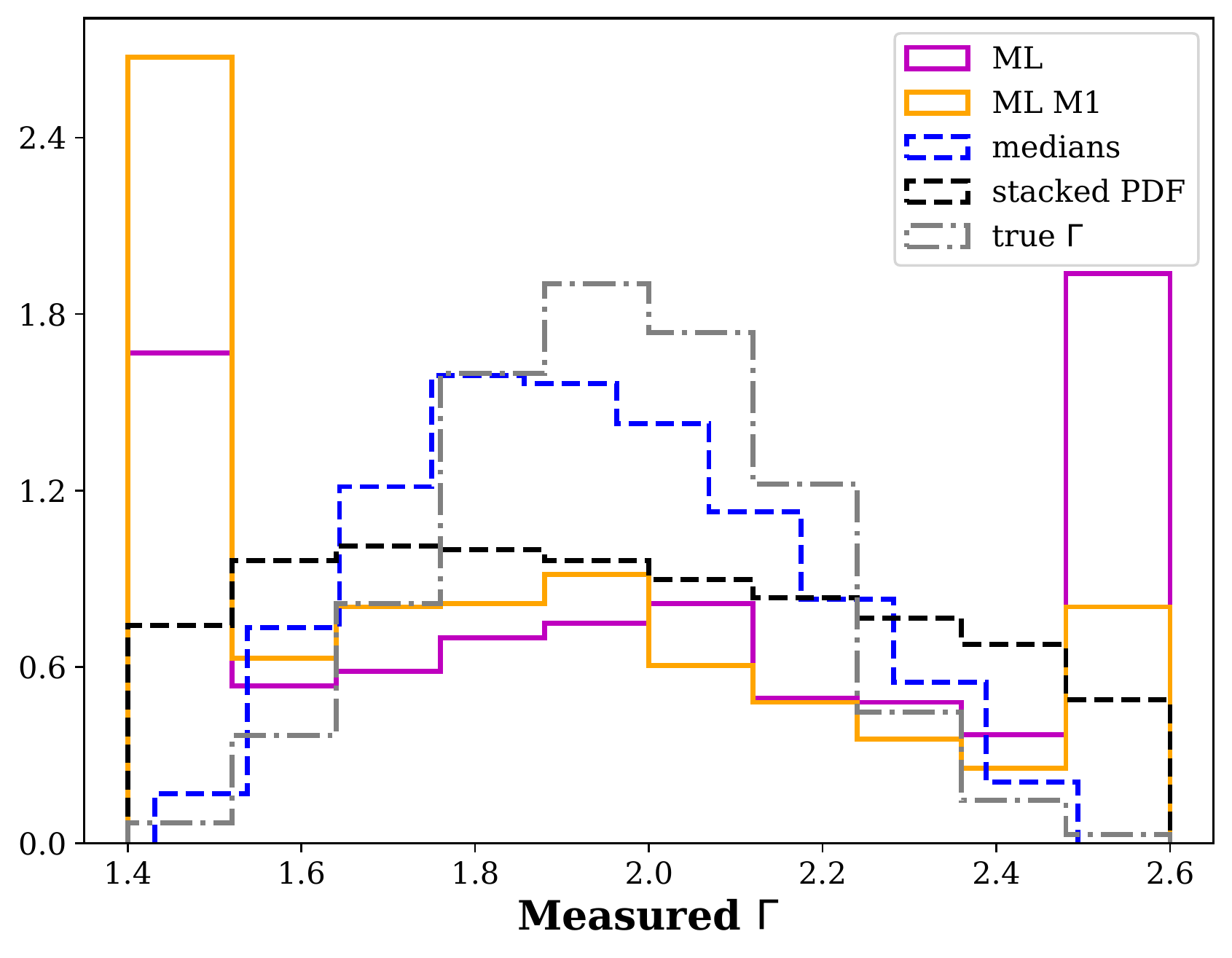}
\caption{Distributions of $\Gamma$ in the absorbed subsample ($N_\mathrm{H}\,>\,10^{22}$\,cm$^{-2}$). The result with the ML approach using the AICc selection is shown by the solid magenta line, and the result using M1 alone is shown by the solid orange line. The results with the Bayesian method using medians and stacked PDF are shown by the dashed blue and black lines, respectively. The gray dot-dashed line shows the parent distribution.}
\label{fig:simul_density_compare}
\end{figure}

\begin{figure*}[tb!]
\centering
\includegraphics[width=17cm]{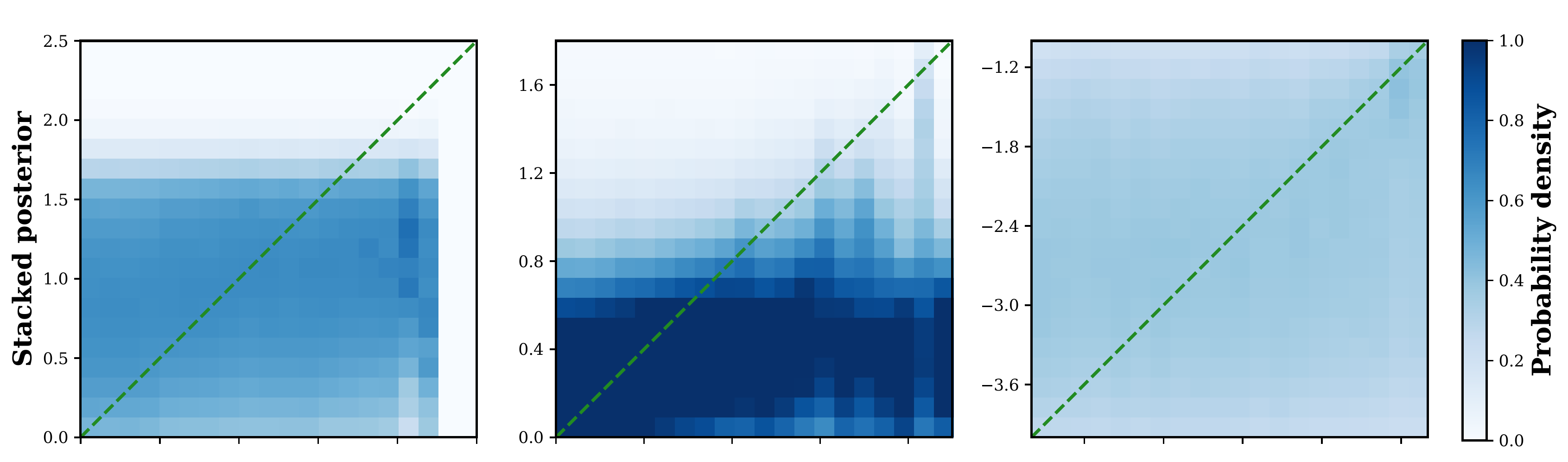}
\includegraphics[width=17cm]{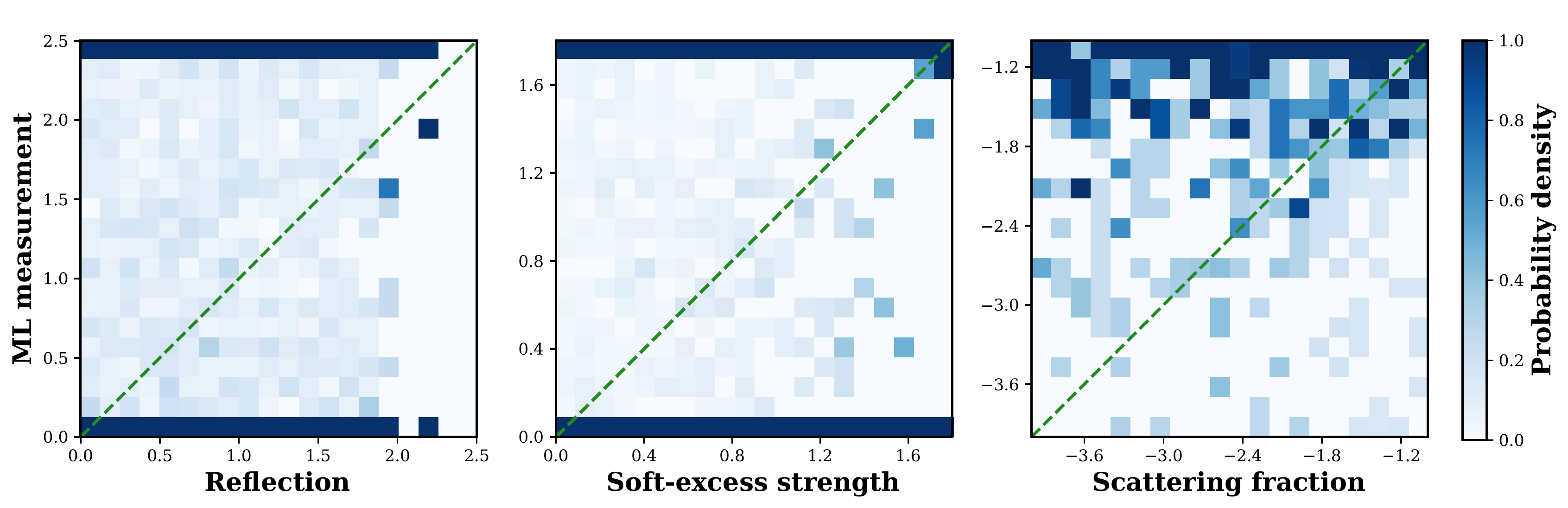}
\caption{Two-dimensional stacked posteriors of secondary spectral parameters, the reflection $R$, soft excess strength $q$, and scattering fraction $f_{scatter}$, which are shown from left to right. The top row shows the results of the Bayesian method, and the bottom row shows those of the ML method. When an ML model does not include one of these secondary components, the corresponding parameter is fixed to 0. The dashed green line shows the reference where the fitted value equals the true one.}
\label{fig:nuisance}
\end{figure*}

\subsubsection{Hydrogen column density $N_\mathrm{H}$}

The results of the fitting of $N_\mathrm{H}$ are shown in the first row of Fig.~\ref{fig:simul_density}. An obvious characteristic shared by both methods is that $N_\mathrm{H}$ is very well constrained for non-CT absorbed sources, defined as $N_\mathrm{H}$ between $10^{22}$ and $10^{24}$\,cm$^{-2}$, while it is poorly constrained for unabsorbed sources. This is mainly due to the loss of sensitivity of the soft X-ray spectrum to the change of $N_\mathrm{H}$ when $N_\mathrm{H}$ becomes too small. Although we are not able to provide a good estimate of the actual $N_\mathrm{H}$ for unabsorbed sources, we can still say with high confidence that they are unabsorbed. In the CT regime, the increased noise and bias is due to the limited information present in the \emph{XMM-Newton} spectra of these objects.

For non-CT absorbed sources, the results of the Bayesian method and ML with selection are comparable in terms of the average bias and dispersion, which are both very small. The measurements of ML using M1 alone, which is the simplest model, are also quite good, but $N_\mathrm{H}$ is biased toward lower values. The reason is that M1 ignores soft excess, thermal emission, and the scattered component, which is compensated for by the smaller absorption of the fit. As shown in the middle left panel of Fig.~\ref{fig:simul_stat}, the Bayesian approach underestimates $N_\mathrm{H}$ from $10^{22}$ to $10^{22.5}$\,cm$^{-2}$ while it overestimates it from $10^{22.5}$ to $10^{23.5}$\,cm$^{-2}$. The reason is that the likelihood changes very little for either very low values or for very high values of $N_\mathrm{H}$, therefore the posterior extends more to the lower or higher limit of $N_\mathrm{H}$. This systematic bias is small, however. 

$N_\mathrm{H}$ of CT sources is not well determined by any method. However, Fig.~\ref{fig:simul_density} shows that the median line of the stacked PDFs remains around $10^{24}$\,cm$^{-2}$, while many results of ML methods are highly biased to much lower values. This shows that our Bayesian method is able to constrain most of the CT sources to lie in the vicinity of the CT regime, while the ML methods quickly lose their constraining power of $N_\mathrm{H}$.

 Although neither the ML nor the Bayesian methods are able to constrain $N_\mathrm{H}$ very well, there are still significant differences in the results for unabsorbed sources. For ML with M1 alone, the vast majority of sources are accumulating at the lower boundary, and the others are randomly scattered up to about $10^{22.5}$\,cm$^{-2}$, which is caused by the fact that we ignored the secondary components. The average bias for ML with selection appears to be largely reduced, but this is just an average effect of two clusters of the fits, one hitting the lower boundary and the other accumulating around $10^{22} \mathrm{cm}^{-2}$. A few sources are also fit with very large $N_\mathrm{H}$ up to $10^{24}$\,cm$^{-2}$. In reality, very few ML fits provide a correct value for $N_\mathrm{H}$ when $N_\mathrm{H}<10^{22}$\,cm$^{-2}$. The reason is that more complex models are selected with our model selection approach, which introduces degeneracies between $N_\mathrm{H}$ and the other parameters. By contrast, the stacked $N_\mathrm{H}$ posterior given by the Bayesian analysis is mostly flat and concentrated below $10^{22}$\,cm$^{-2}$. This means that even though $N_\mathrm{H}$ is not well constrained, the classification of the sources between the absorbed and unabsorbed categories is accurate. The loss of sensitivity of the data to $N_\mathrm{H}$ blurs the posterior, but does not introduce a strong bias. 

To show the different performances of the approaches on unabsorbed sources better, we plot the measured $N_\mathrm{H}$ distributions of the unabsorbed subsample in Fig.~\ref{fig:nh_unabsorbed}. It is clear that both ML approaches fail at reproducing the expected $N_\mathrm{H}$ distribution, which is flat. Very few sources with $N_\mathrm{H}$ between $10^{20}$ to $10^{21}$\,cm$^{-2}$ are recovered by ML, and the false impression arises that this distribution is bimodal. On the other hand, our Bayesian method returns mostly flat PDFs in this range, which mostly follow the prior. However, even though the $N_\mathrm{H}$ of individual sources is not well constrained, the PDFs still contain some information, for instance, the maximum allowed $N_\mathrm{H}$, unlike the ML results, for which a large fraction of sources hit the lower boundary of the allowed range. The method that we devised to infer the properties of the parent population from the output PDFs is presented in Paper~\rom{2}.

\subsubsection{Spectral index $\Gamma$}

The results of the fitting of the photon index $\Gamma$ using the two methods are shown in the second row of Fig.~\ref{fig:simul_density}. There are clearly significant biases in the ML fit, where the fitted photon indices for many sources hit the fit boundaries at $\Gamma=1.4$ and $2.6$, and the formal error on $\Gamma$ is very small and largely underestimated. In addition, the photon indices obtained with ML with model selection show a constant bias toward larger $\Gamma$, while the fits for ML with M1 alone show a constant negative bias. The negative bias when M1 alone is applied is mostly caused by the presence of the reflection component, which is ignored by the simplest model, while the positive bias of the other is due to the overfitting of secondary components, for instance, fitting the reflection parameter to the maximum value (see Fig.~\ref{fig:nuisance}).

By contrast, the Bayesian method shows a more reasonable systematic bias and very few outliers. In the range of 1.8 to 2.2 where most of the photon indices are expected \citep{Nandra1994,Buchner2014}, the Bayesian method has a very small bias and dispersion, as shown in the middle column of Fig.~\ref{fig:simul_stat}. The bias becomes larger when $\Gamma$  approaches the boundaries, mainly because the edges of the prior (see Sect.~\ref{model priors}) cut the low- or high-$\Gamma$ extension of the posteriors. In addition, only a few sources lie in this range in our simulation because we drew our true $\Gamma$ from a Gaussian distribution centered at 1.95 (see Sect.~\ref{simulated sample}). The results can therefore be affected by small-number statistics. These values are not really expected in reality, however \citep{Buchner2014}. If the true $\Gamma$ really follows the distribution of \citet{Buchner2014}, our method would be able to recover $\Gamma$ for the vast majority of the objects with little bias and dispersion. The average bias of the ML methods is very misleading. The average bias of the ML methods with model selection for sources with true $\Gamma$ around 2 would indeed be very small, even though most of the measurements are very poor, as shown in Fig.~\ref{fig:simul_density}, and they are randomly scattered to the two boundaries, so that the average value happens to be very close to the true value. The dispersion is also meaningless in these cases because it mostly depends on where the boundaries of the fits are defined.

Because $N_\mathrm{H}$ strongly affects the measurement of $\Gamma$, we divided our simulated sample into absorbed and unabsorbed subsamples with the boundary set at $N_\mathrm{H} = 10^{22}$\,cm$^{-2}$. We plot the results of ML with model selection and our Bayesian method in Fig.~\ref{fig:simul_density_nh}. The constraints on $\Gamma$ with both methods are much better for unabsorbed sources than for absorbed sources, as expected. However, there are still some catastrophic failures of ML even in the unabsorbed subsample, while the Bayesian method gives smooth stacked PDFs with very small bias and dispersion. For the absorbed sub-sample, both approaches are unable to measure $\Gamma$ correctly, but unlike our Bayesian method, which gives smooth PDFs, ML severely biases the measurements of $\Gamma$ to extreme values, often reaching the fit boundaries. Again, the median of ML is misleading because very few measurements of $\Gamma$ using ML are meaningful (i.e., do not hit the fit boundaries). As also shown in Fig.~\ref{fig:simul_density_compare}, more than 40\% of the absorbed sources are fit by ML with $\Gamma$ around 1.4 or 2.6. By contrast, in the Bayesian approach, the medians of the $\Gamma$ posteriors roughly recover a correct $\Gamma$ distribution, which in our case is a Gaussian around 1.95 (see Sect.~\ref{simulated sample}). The posteriors are close to the flat prior from 1.5 to 2.5, however, which can be seen from the stacked PDF. With a flat posterior the median of the posterior is equal to the central value of the range, which is arbitrary, and in fact, neither the medians nor the stacked PDF provides a correct reconstruction of the parent population; we address this point in Paper~\rom{2}. This also shows that when there is little information in the data to constrain a parameter, ML is very unstable and tends to adopt highly biased values, while in the Bayesian approach, the values of the parameter remain constrained by the prior to a range of acceptable values according to our previous knowledge.

\subsubsection{X-ray luminosity $L_\mathrm{X}$}
We compared the intrinsic 2--10\,keV $L_\mathrm{X}$ recovered with the different fitting approaches with the true $L_\mathrm{X}$. In the case of the ML methods, $L_\mathrm{X}$ was computed using the best-fit parameters. In the Bayesian analysis, a PDF of $L_\mathrm{X}$ was obtained by randomly drawing many sets of spectral parameters from the corresponding posteriors and then computing $L_\mathrm{X}$ for each of them.

Fig.~\ref{fig:simul_density} shows that all methods are able to constrain $L_\mathrm{X}$ well, but there are still significant differences in their performances. First of all, for some sources the ML method with model selection biases $L_\mathrm{X}$ to much higher values, some as high as $10^{46}$\,erg\,s$^{-1}$ , while the true $L_\mathrm{X}$ is only about $10^{43}$\,erg\,s$^{-1}$. Even though these outliers only amount to about 3\% of the total sample, the biased measurements of $L_\mathrm{X}$ might affect the computation of the luminosity function at the bright end strongly because there are intrinsically few very luminous sources in the sample. This bias is mostly due to the degeneracy between unabsorbed power law and absorbed power law plus scattering: if the source is intrinsically faint, we can fit it with a source with a much higher luminous, but highly absorbed, together with a scattering component. The scattering, although its luminosity is lower than 10\% of that of the primary power law, is still enough to compensate for the soft X-ray part of the spectrum. When the ML method is used with M1 alone, it appears to be able to measure $L_\mathrm{X}$ with a small bias and only a few outliers, even though its measurements of $N_\mathrm{H}$ and $\Gamma$ are often strongly biased. However, this is simply because in the 2--10\,keV band, where $N_\mathrm{H}$ has a smaller impact, a simple power law with harder photon index can be still a good fit of a spectrum, which is intrinsically a softer power law plus reflection.

By contrast, the Bayesian approach is able to constrain $L_\mathrm{X}$ throughout most of the range, without obvious bias and dispersion. The degeneracy observed with the ML method is largely removed by the use of strong priors on secondary components. This shows that, on one hand, the measurement of the hard-band $L_\mathrm{X}$ is mostly robust, even when the column densities and photon indices are not measured accurately; on the other hand, the Bayesian method is still more robust than the ML one, with or without model selection.

\subsubsection{Secondary parameters}

The typical S/N in our simulated sample does not allow us to constrain the secondary spectral parameters such as reflection and soft excess. Fig.~\ref{fig:nuisance} shows the corresponding posteriors, which simply follow their priors, meaning that we do not gain any information from the data for these parameters of most sources. We also show the ML measurements of secondary parameters in Fig.~\ref{fig:nuisance}, except for those of the thermal component because they are most often ignored in the ML analysis (see Sect.~\ref{ML approach}).

Although both methods are unable to constrain these parameters, there are still significant differences in the results. The ML approach results in extreme values for most of the sources. For example, $R$ is mostly fit with either the maximum value 2.5 or 0 (ignored), even though the values at the boundaries are unlikely to exist in reality. These results give us incorrect information on the parameters and bias the measurements of the other parameters. On the other hand, the stacked PDFs of the Bayesian method are smoothly distributed in the areas that are constrained by the priors. This clearly shows that the data are not able to constrain these parameters, but their errors are correctly propagated into the posteriors of the main parameters.

\section{Discussion}
\label{Discussion}

\subsection{Bayesian versus ML approaches}

The ML approach shares some common features with the Bayesian approach in the sense that they both use the same likelihood function as an essential ingredient to estimate their parameters. However, there are also fundamental differences between the two methods that strongly impact the results when they are applied them to a large number of AGN X-ray spectra with low to moderate data quality, as is typical of surveys. 

The first difference is that the ML approach does not allow the use of priors to place constraints on secondary parameters, so that fitting complex models to data with a low S/N results in very unstable ML parameters. As we have shown in Fig.~\ref{fig:simul_density_compare} and Fig.~\ref{fig:nuisance}, when the data contain little information on parameters such as $R$, the value recovered by the ML approach tends to hit the boundaries of the allowed range. When this approach is applied blindly, it might naively be thought that the parameters are well constrained. Because some of the parameters of the model are correlated, these highly biased measurements can affect those of some of the parameters of interest ($N_\mathrm{H}$, $\Gamma$, $L_\mathrm{X}$).

The ML M1 and ML methods that we have presented mimic the approach followed in most of the literature when Bayesian methods are not used. The first approach, ML M1, ignores secondary components, which can bias some spectra that appear very simple but are actually complex (see Sect.~\ref{bias example}), due to the degeneracy between the primary and secondary parameters. The application to the simulated sample shows that ML M1 biases $N_\mathrm{H}$ toward lower values and quickly loses its constraining power when the absorption is lower than $10^{22}$\,cm$^{-2}$ (see Fig.~\ref{fig:simul_density}). It also constantly biases $\Gamma$ to harder values, as shown in Fig.~\ref{fig:simul_stat}. Therefore the simple model is not able to recover sensible parameters even though the fit appears reasonable, so that more complex models are required.

We tested an objective and automatic model selection procedure based on six models with increasing complexity. Model selection was performed using the AICc estimator. Somewhat surprisingly, the simplest model is always excluded. Instead, moderately complex models, most often with reflection (M3), but sometimes with soft excess (M2), are preferred. \citet{Buchner2014} also found that reflection is usually needed (see details in Sect.~\ref{discussion of comparison}). The overall improvement of the fits is demonstrated in Fig.~\ref{fig:ml_cstat} with approximate estimators of GoF that are largely reduced after model selection. $N_\mathrm{H}$ is relatively well measured for non-CT absorbed sources with smaller bias than when M1 is used alone. However, the measured $N_\mathrm{H}$ for unabsorbed sources are still unusable, as shown in Fig.~\ref{fig:nh_unabsorbed}, and the ML method with model selection is still unable to provide a reliable estimate of the photon index $\Gamma$, as most of the fitted values hit the upper and lower boundaries of the allowed range (see Fig.~\ref{fig:simul_density} and \ref{fig:simul_density_compare}). While all methods (including M1) can accurately retrieve the intrinsic luminosity of the source, when we account for about 3\% of the sources in our sample, there is a small cluster of sources with highly biased $L_{\mathrm{X}}$, which is probably due to the strong degeneracy with some secondary parameters (see Fig.~\ref{fig:nuisance}). Therefore it is still not sufficient to apply ML with more complex models to constrain the main spectral parameters without obvious bias.

By contrast, our Bayesian approach is able to apply the full model (M6) to all the sources and still obtain mostly unbiased fits for the main parameters of interest, namely $N_\mathrm{H}$, $\Gamma$ , and $L_{\mathrm{X}}$. The main reason is that the Bayesian approach uses informative priors to constrain secondary parameters, avoiding the impact of secondary components being fitted with unrealistically extreme values, as discussed above. In the end, it marginalizes over the secondary parameters that are unconstrained, as well as over the nuisance parameters of the background models, to obtain the PDFs of the main parameters. This propagates the uncertainties correctly. Another reason is that our Bayesian approach samples the entire parameter space with \texttt{MultiNest}, while ML approaches are prone to ending up in local maxima, especially when the models are very complex and the parameter space is highly dimensional.

Another main difference is that the Bayesian approach produces PDFs that contain the full posterior probability information, while ML only provides point estimates. In some case, the PDFs may still be biased, for instance, those of $N_\mathrm{H}$ for unabsorbed sources. However, even when they are biased, the PDFs practically always cover the true values, so that it can still help to reproduce the parent parameter distributions, which is the main goal of this analysis. In contrast, the ML approach leads to many catastrophic failures, in which case it is impossible to reconstruct the parent population, for example, in the two cases shown in Fig.~\ref{fig:nh_unabsorbed} and \ref{fig:simul_density_compare}. In Paper~\rom{2} we propose a method for recovering the parent distribution in the presence of biased PDFs.

\subsection{Comparison with other Bayesian methods}
\label{discussion of comparison}

\begin{figure*}[h!]
\centering
\includegraphics[width=17cm]{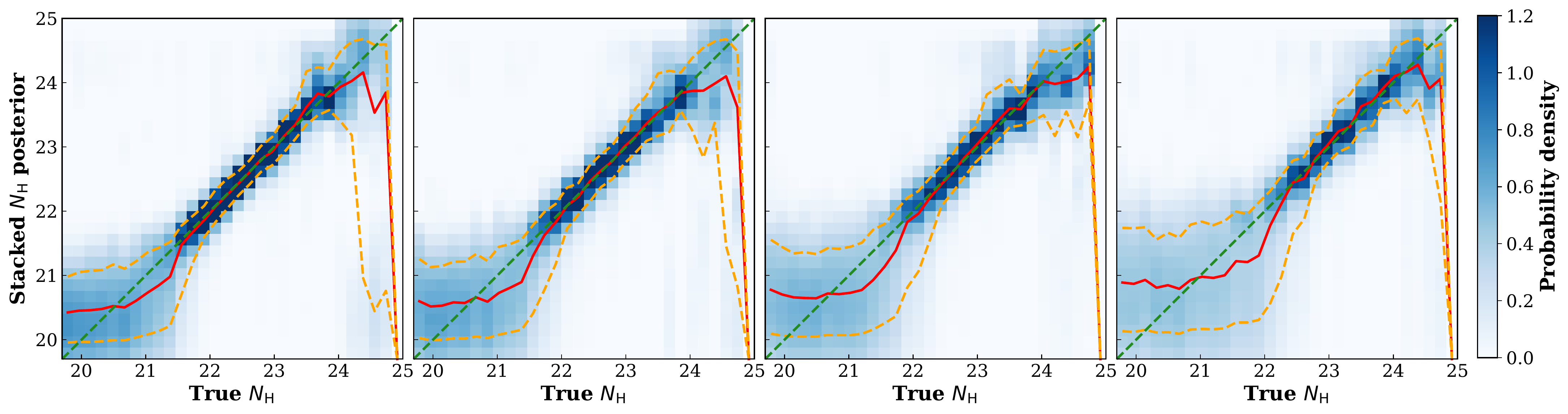}
\caption{Two-dimensional stacked PDFs of $N_\mathrm{H}$ of subsamples in different redshift bins: [0,0.6], [0.6,1.0], [1.0,1.5], and [1.5,4], from left to right. The red line shows the medians of the distribution in each bin, and the dashed orange lines are the 16\% and 84\% percentiles. The dashed green line shows the reference where the fitted value equals the true one.}
\label{fig:nh_evol}
\end{figure*}

Three groups have already pioneered Bayesian analysis of AGN X-ray spectra: \citet{Buchner2014}, \citet{Fotopoulou2016}, and \citet{Ruiz2020}. A common feature of the spectral analysis methods in these works, as well as ours, is that a nested-sampling algorithm such as \texttt{MultiNest} is applied to sample the full parameter space.

\citet{Fotopoulou2016} adopted a similar but much simpler approach to ours to fit the XXL-1000-AGN sample. The background spectra and source plus background were fit simultaneously in order to preserve the Poisson statistics. The background spectra were modeled with the same XB model as we used. However, the NXB was modeled with a combination of several emission lines and continuum components, which are phenomenologically represented by blackbodies and/or power laws, while we used the FWC data as described in Sect.~\ref{nxb model}. The most important difference between our method and that of \citet{Fotopoulou2016} is that they adopted a simple absorbed power-law model for all the sources. Even though noninformative priors were used for $\Gamma$ and $N_\mathrm{H}$, ignoring secondary components such as reflection and soft excess could bias the PDFs. For example, we have shown in Sect.~\ref{bias example} that even when the data appear to be simple, the spectrum may be intrinsically complex. 

\citet{Buchner2014} analyzed the X-ray spectra of about 350 AGN in the four Ms \textit{Chandra} Deep Field South using a Bayesian method to select among ten physically motivated models with different complexities, from a simple power law to a complex model including secondary components. The components considered in the models include the primary power law, absorption, reflection, and scattering. Soft excess was neglected because it lies outside of the observed energy range of most of the sources because of the redshift. Background and source plus background spectra were fit separately, with the background fit first using a Gaussian mixture model and then fixed in the fitting of the source plus background spectra. This is a common approach to model background in general, but our NXB model represents the most advanced modeling of the XMM background \citep{Ghirardini2018}. A Jeffreys prior was also applied to $N_\mathrm{H}$, but a strong Gaussian prior with a mean 1.95 and a standard deviation 0.15 was adopted for $\Gamma$. In general, our simulations have shown that measuring $\Gamma$ is difficult, especially for absorbed sources, thus they only marginalized over it. In addition, noninformative priors were applied on secondary components, while we used informative priors obtained from deep surveys to constrain the complex models. Another difference is that we included the selection function into our likelihood function, in order to exclude undetectable fits in the source detection band.

A further difference between \citet{Buchner2014} and our work is that they employed model selection among many models with different complexities, while we consistently used the most complex model for all the sources. However, their overall best model after model selection ends up with \texttt{powerlaw+torus+pexmon+scattering}, where powerlaw is the primary emission, \texttt{torus} is the absorber, \texttt{pexmon} is the reflection, and the last item is the scattering. This is also a complex model and very similar to ours, especially considering that soft excess and thermal emission do not apply to their mostly high-z sources. For the low-z universe, however, soft excess and thermal emission are present in the soft X-ray band. General criteria are used in \citet{Buchner2014} for model selection, such as the Bayes factor and BIC and AIC values. We propose in this paper a generalization that is valid for sources at all redshifts, more advanced NXB background modeling for \emph{XMM-Newton,} and, most importantly, the use of informative priors for the secondary components, as well as a complete likelihood that takes the selection function into account. Similar approaches were used in \citet{Liu2016} and \citet{Liu2021} to analyze the point-like X-ray sources in the northern field of the \emph{XMM}-XXL survey and those in the eROSITA final equatorial-depth survey (eFEDS), respectively, except that \citet{Liu2016} applied the best model of \citet{Buchner2014} for all the sources, while \citet{Liu2021} applied simpler models, including the primary power law, with $N_\mathrm{H}$ or $\Gamma$ fixed in some cases, and sometimes a soft-excess component.

\citet{Ruiz2020} also used a similar approach to \citet{Buchner2014} to analyze about 22'700 sources in the 3XMM catalogue to search for obscured sources. Four phenomenological models were considered, including an absorbed power-law model, an absorbed thermal model, an absorbed thermal plus power-law model, and an absorbed double power-law model. Other secondary components, such as reflection and soft excess, were neglected. Similarly to \citet{Buchner2014}, noninformative priors were applied on secondary components, which makes it more difficult to constrain complex models. The best model was selected using the Kolmogorov-Smirnov (KS) statistics and the corresponding $p$-value, and the evidence provided by \texttt{MultiNest} for each model. As a result, the simple absorbed power-law model was accepted or preferred for most the sample (90\%). Another difference is that \citet{Ruiz2020} used the W statistics in \texttt{xspec} to take the Poisson background into account, but this is not exact and will lead to biases for weak sources and small numbers of counts in the background spectrum (see Appendix B of the Xspec User Guide\footnote{\url{https://heasarc.gsfc.nasa.gov/xanadu/xspec/manual/XSappendixStatistics.html}}). In our work, we modeled the source plus background and background simultaneously, fully preserving the Poisson statistics.

A final difference in the different approaches is that the whole fitting process is done outside of the usual X-ray spectral fitting softwares such as \texttt{xspec}. All spectra are binned in the same way and extracted in a way to have a single response (for each instrument), and models are preconvolved with the response once for all in large model matrices. This results in a large gain in the computing efficiency of the spectral analysis.

\subsection{Effect of the redshift on $N_\mathrm{H}$}

The constraint on $N_\mathrm{H}$ is highly dependent on the redshift, as shown in Fig.~\ref{fig:nh_evol}. We divided the total sample into four redshift bins: [0, 0.6], [0.6, 1.0], [1.0, 1.5], and [1.5, 4], so that the number of sources in each bin is roughly the same. In the first redshift bin, there is a very clear transition point around $10^{21.5}$\,cm$^{-2}$, above which the measurements of $N_\mathrm{H}$ are very well constrained, while the measurements fail below it and the posterior becomes uniformly redistributed, but still remaining in the unabsorbed region. As redshift increases, this transition point moves towards higher absorption. Despite the increased noise in the plot of the highest redshift bin due to the decreased S/N on average, a transition point can be clearly identified above $10^{22}$\,cm$^{-2}$. The reason is that the information for $N_\mathrm{H} \sim 10^{22}$\,cm$^{-2}$ is concentrated below about 2\,keV in the rest frame, and that the redshift moves the informative part of the spectrum to lower energies outside of the considered spectral range.  

There is also a similar effect for CT sources. In the first redshift bin, the uncertainties in $N_\mathrm{H}$ increase very rapidly above $10^{24}$\,cm$^{-2}$. However, at higher redshifts, the uncertainty in the $N_\mathrm{H}$ of CT sources becomes clearly smaller, with more probability constrained to the CT regime and its vicinity. This can be also naturally explained by the fact that the absorption feature and Compton hump are redshifted to lower energies, which gives the soft X-ray spectrum stronger constraining power over $N_\mathrm{H}$.

\subsection{Effect of uncertainties in the redshift}

The redshift is obviously also an important parameter in the fitting of AGN X-ray spectra in large surveys. Because it is time consuming to obtain spectroscopic redshifts (spec-z) for each source, photometric redshifts (photo-z) must be used for some of the sources. Many photo-z codes today produce probability distribution function (PDZ) using template fitting \citep{Benitez2000} or machine learning \citep{Carrascokind2013}, as well as many other subsequent works.

In principle, the PDZs can be used as a prior on a redshift parameter that we can fit using our approach without any modification, and in fact, in the model presented above, the redshift is already a free model parameter with a very strong prior. In the end, the photo-z is marginalized to propagate its uncertainty into the posterior distribution of the main parameters, similarly to the secondary and nuisance parameters. However, considering that our model is already very complicated and the fit of soft X-ray spectrum is quite sensitive to redshift, especially for low-S/N data, in this paper we assumed that a spec-z, or an excellent quality photo-z, is available, such that the prior on the redshift is effectively a Dirac distribution. In addition, photometric redshifts are subject to catastrophic failure, such that the PDZs do not necessarily represent the true posterior probabilities of the redshift. Thus we consider that including the full PDZs, while technically perfectly feasible within our analysis framework, would uncover problems that are not directly related to the analysis of X-ray data, and thus are beyond the scope of this work.

\section{Conclusion}
\label{Conclusion}

We developed a novel Bayesian method using nested sampling to automatically fit a large number of AGN X-ray spectra at low-S/N level. A single physically motivated model was adopted for all the sources, including the primary power law, soft excess, reflection, scattering, and emission from thermal plasma. We applied noninformative priors for the most relevant spectral parameters, for instance, $N_\mathrm{H}$ and $\Gamma$, aiming at constraining their distributions from the data, and informative priors derived from deep observations on the other parameters that are difficult to constrain with low-S/N data, to properly propagate their uncertainties by marginalizing over them. We implemented the survey selection function directly into our Bayesian model by modifying the likelihood to discard sources whose count rates in the detection band were below the detection threshold. This complete likelihood was used to penalize undetectable solutions due to the difference between the source detection band and the spectral fitting band. We modeled source and background spectra simultaneously to preserve Poisson statistics, and the NXB model was derived with filter-wheel-closed data. The spectra were extracted from aperture photometry with consistent energy binning, which allowed us to tabulate \texttt{xspec} models by convolving the theoretical models with the constant redistribution matrix, largely speeding up both the spectral extraction and model fitting. 

We tested our method on a realistic sample of 5000 simulated spectra (plus about 500 CT objects), most of which have low S/N (3--5). Based on the results in Sect.~\ref{simulation}, our Bayesian method is able to constrain the main spectral parameters, namely $N_\mathrm{H}$ and $\Gamma$, and to recover the intrinsic X-ray luminosity even at low-S/N level. The $N_\mathrm{H}$s of absorbed sources are accurately recovered, and for unabsorbed sources, we obtain a correct upper limit of $N_\mathrm{H}$. $\Gamma$ was fit with relatively large uncertainties for absorbed sources, but for unabsorbed sources, we were able to obtain a good constraint. The intrinsic X-ray luminosity is very well recovered for the whole sample, which is important for studying the demographics of AGN population. On the other hand, parameters of secondary components such as reflection and soft excess are poorly constrained, but this is expected. Applying informative priors allowed us to marginalize over the range of acceptable model parameters and to propagate the uncertainties to the parameters of interest.

The advantages of our new approach of AGN X-ray spectral fitting can be summarized as follows:

\begin{itemize}

\item Consistent physical model: A comprehensive physics-driven model was applied to all sources without model selection and regardless of the S/N. It overcomes the bias introduced by selecting models based on the apparent model complexity.

\item Background modeling: Source and background spectra were modeled simultaneously to preserve the Poisson statistics. The background was modeled using a theoretical model for XB and normalized filter-wheel-closed data for NXB.
\item Automatic and fast analysis: The whole spectral analysis process is fully automatic from spectral extraction to parameter estimation, and it is computationally very efficient based on a pure Python code that bypasses the need to use software such as \texttt{xspec}. This is a key requirement of large X-ray surveys.

\item Bayesian parameter estimation: The Bayesian analysis using nested sampling allowed us to fully explore the high dimensional parameter space and to obtain the full posterior distribution of each parameter. The uncertainties of secondary and nuisance parameters were correctly taken into account by marginalizing over them, which gives us a mostly unbiased estimation of parameters of interest, such as $N_\mathrm{H}$, $\Gamma$ and $L_{\mathrm{X}}$.

\end{itemize}

In Paper~\rom{2} we expand the Bayesian spectral analysis to determine the properties of the parent population using a fully Bayesian approach. We take full advantage of the posteriors of the parameters  derived using the method proposed here.

\begin{acknowledgements}
      L.G. acknowledges partial support from the Swiss National Science Foundation.
\end{acknowledgements}

\bibliographystyle{aa}
\bibliography{ref.bib}


\end{document}